\documentclass{ifacconf}%
\usepackage{graphicx}
\usepackage{natbib}
\usepackage{graphics}
\usepackage{epsfig}
\usepackage{mathptmx}
\usepackage{times}
\usepackage{amsmath}
\usepackage{amssymb}
\usepackage{amsfonts}
\usepackage{graphicx}
\usepackage{titlesec}
\usepackage{xcolor}%
\providecommand{\U}[1]{\protect\rule{.1in}{.1in}}
\begin{document}
\begin{frontmatter}
\title{Stochastic Flow Models with Delays and Applications to
	Multi-Intersection Traffic Light Control \thanksref{footnoteinfo} }
\thanks[footnoteinfo]{Supported in part by NSF under grants ECCS-1509084, CNS-1645681, and IIP-1430145, by AFOSR under grant FA9550-15-1-0471, by the DOE under grant DE-AR0000796, by the MathWorks and by Bosch.}
\author[First]{Rui Chen}
\author[First]{ Christos G. Cassandras}
\address[First]{Division of Systems Engineering, Boston University, Brookline, MA, USA, (e-mail: ruic@bu.edu, cgc@bu.edu).}

\begin{abstract}                
We extend Stochastic Flow Models (SFMs), used for a large class of discrete event and
hybrid systems, by including the delays which typically arise in flow
movement. We apply this framework to the multi-intersection traffic light
control problem by including transit delays for vehicles moving from one
intersection to the next. Using Infinitesimal Perturbation Analysis (IPA) for
this SFM with delays, we derive new on-line gradient estimates of several
congestion cost metrics with respect to the controllable green and red cycle
lengths. The IPA estimators are used to iteratively adjust light cycle lengths
to improve performance and, in conjunction with a standard gradient-based
algorithm, to obtain optimal values which adapt to changing traffic
conditions. We introduce two new cost metrics to better capture congestion and
show that the inclusion of delays in our analysis leads to improved
performance relative to models that ignore delays.
\end{abstract}
\end{frontmatter}

\section{INTRODUCTION}

Stochastic Flow Models (SFMs) capture the dynamic behavior of a large class of
hybrid systems (see \cite{cassandras2009introduction}). In addition, they are
used as abstractions of Discrete Event Systems (DES), for example when
discrete entities accessing resources are treated as flows. The basic building
block in a SFM is a queue (buffer) whose fluid content is dependent on
incoming and outgoing flows which may be controllable. By connecting such
building blocks together, one can generate stochastic flow networks which are
encountered in application areas such as manufacturing systems
(\cite{armony2015patient}), chemical processes (\cite{yin2013data}), water
resources (\cite{anderson2015applied}), communication networks
(\cite{CSP01TAC}) and transportation systems (\cite{geng2015multi}). Figure
\ref{twonodeSFM} shows a two-node SFM, in which an on-off switch controls the
outgoing flow for each node. When the switch at the output of node $1$ is
turned on, a \textquotedblleft flow burst\textquotedblright\ is generated to
join the downstream node $2$. Flow models commonly assume that this flow burst
can instantaneously join the downstream queue, thus ignoring potentially
significant delays before this can happen. Incorporating such delays through
more accurate modeling is challenging but crucial in better evaluating the
performance of the underlying system and seeking ways to improve it.

Control mechanisms used in SFMs often involve gradient-based methods in which
the controller uses estimates of the performance metric sensitivities with
respect to controllable parameters in order to adjust the values of these
parameters and improve (ideally, optimize) performance. Infinitesimal
Perturbation Analysis (IPA) is a method of general applicability to stochastic
hybrid systems (see \cite{Cassandras2010},\cite{Wardi2010}) through which
gradients of performance measures may be estimated with respect to several
controllable parameters based on directly observable data. The applications of
IPA and its advantages have been reported elsewhere (e.g.,
\cite{Cassandras2010},\cite{fleck2016adaptive}) and are summarized here as
follows: $(i)$ IPA estimates have been shown to be unbiased under very mild
conditions (\cite{Cassandras2010}). $(ii)$ IPA estimators are robust with
respect to the stochastic processes involved. $(iii)$ IPA is
event-driven, hence scalable in the number of events in the system, not the
(much larger) state space dimensionality. $(iv)$ IPA possesses a
decomposability property (\cite{YaoCas2011}), i.e., IPA state derivatives
become memoryless after certain events take place. $(v)$ The IPA methodology
can be easily implemented on line, allowing us to take advantage of directly
observed data.

While IPA has been extensively used in SFMs, the effect of delays between
adjacent nodes, as described above, has not been studied to date. Thus, the
contribution of this paper is to incorporate delays in the flow bursts that
are created by on-off switching control (see Fig. \ref{twonodeSFM}) into the
standard SFM and to develop the necessary extensions to IPA for such systems.
In addition, an application of SFMs with delays to the Traffic Light Control
(TLC) problem in transportation networks is included.

The rest of the paper is organized as follows. In Section 2, we extend the
standard multi-node SFM to include delays. In Section 3 we adapt this model to
the TLC problem by explicitly modeling the delay experienced by vehicles
moving from one intersection to the next. This allows us to introduce two new
cost metrics for congestion that incorporate the effect of delays. In Section
4, we carry out IPA for the TLC problem and in Section 5 we provide simulation
examples comparing performance results between a model considering traffic
delays and one which does not, showing that the former achieves improved performance.

\section{STOCHASTIC FLOW MODELS WITH DELAYS}

\begin{figure}[pt]
	\label{twonodeSFM} \centering
	\includegraphics[scale=0.4]{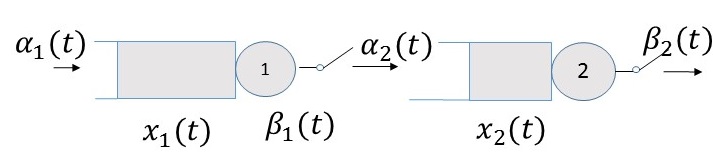}\caption{A two-node SFM.}%
\end{figure}

Consider a two-node SFM as in Fig. \ref{twonodeSFM} and let $\{\alpha
_{i}(t)\}$ and $\{\beta_{i}(t)\}$, $i=1,2$, be the \emph{incoming flow }and
\emph{outgoing flow processes }respectively. We emphasize that these are both
treated as random processes. We define $x(t)=[x_{1}%
(t),x_{2}(t)]$, where $x_{i}(t)\in\mathbb{R}^{+}$ is the flow content of node
$i$ (we assume that all variables are left-continuous.) The dynamics of this
SFM are
\begin{equation}
\dot{x}_{i}(t)=\left\{  \vspace*{0pt}%
\begin{array}
[c]{l}%
0\\
\\
\alpha_{i}(t)-\beta_{i}(t)
\end{array}
\right.
\begin{array}
[c]{l}%
\text{if }x_{i}(t)=0,\text{ }\alpha_{i}(t)\leq\beta_{i}(t)\\
\text{or }x_{i}(t)=c_{i},\text{ }\alpha_{i}(t)\geq\beta_{i}(t)\\
\text{otherwise}%
\end{array}
\label{x_dynamics}%
\end{equation}
where $c_{i}$ is the content capacity of $i$ and $\beta_{i}(t)$ is
\begin{equation}
\vspace*{-1pt}\beta_{i}(t)=\left\{
\begin{array}
[c]{l}%
h_{i}(t)\\
0
\end{array}
\right.
\begin{array}
[c]{l}%
\text{if }G_{i}(t)=1\\
\text{otherwise}%
\end{array}
\label{SFMbeta_dynamics}%
\end{equation}
in which $h_{i}(t)$ is the instantaneous outgoing flow rate at node $i$, and $G_{i}(t)\in\{0,1\}$, $i=1,2$ is a switching controller.
We also
define a clock state variable $z_{i}(t)$ for each switching controller
$G_{i}(t)$:
\begin{align}
\dot{z}_{i}(t) &  =\left\{
\begin{array}
[c]{l}%
1\\
0
\end{array}
\right.
\begin{array}
[c]{l}%
\text{if }G_{i}(t)=1\\
\text{otherwise}%
\end{array}
\label{zi_dynamics}\\
z_{i}(t^{+}) &  =0\text{ \ \ if }G_{i}(t)=1\text{ and }G_{i}(t^{+})=0\nonumber
\end{align}
Thus, when $G_{1}(t)=1,$ $t\in\lbrack t_{1},t_{2})$, $G_{1}(t_{1}^{-})=0$, a
flow burst is created at node $1$ (when $x_{1}(t_{1})>0$). In general, several
such flow bursts may be created over $(t_{1},t_{2}]$, depending on the values
of $\alpha_{1}(t)$, $h_{1}(t),$ $t\in(t_{1},t_{2}]$. In SFMs studied to date, we ignore the delay
incurred by any such flow burst being transferred between nodes and assume
that it instantaneously joins the queue at node $2$. Under this assumption,
\[
\vspace*{-1pt}\alpha_{2}(t)=\left\{
\begin{array}
[c]{l}%
\alpha_{1}(t)\\
\beta_{1}(t)
\end{array}
\right.
\begin{array}
[c]{l}%
\text{if }x_{1}(t)=0,\text{ }\alpha_{1}(t)\leq\beta_{1}(t)\\
\text{otherwise}%
\end{array}
\]
In what follows, we extend the SFM to include the aforementioned delay which
depends on when a flow burst actually joins the downstream queue, an event
that we need to carefully specify. While a flow burst is in transit between
nodes $1$ and $2$, let $x_{12}(t)$ be its size, i.e.,the flow volume in transit before it joins $x_{2}(t)$. For
simplicity, we assume that each flow burst is maintained during this process
(i.e., the burst may not be separated in two or more sub-bursts). We will use
$L$ to denote the physical distance between nodes 1 and 2.

\begin{figure}[ptb]
	\centering
	\includegraphics[scale=0.22]{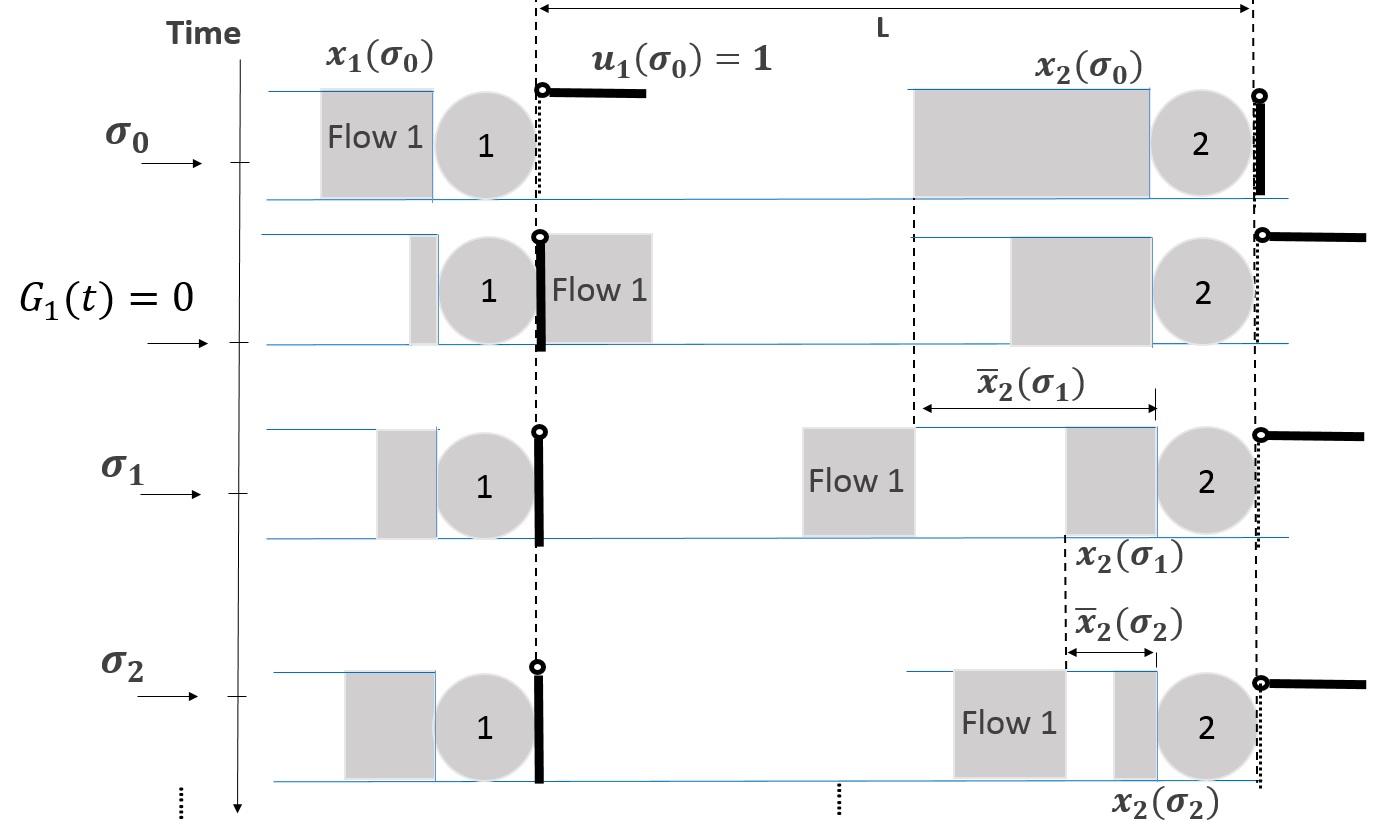} \caption{Typical evolution of a
		flow burst in transit.}%
	\label{SFMwithdelay}%
\end{figure}

Predicting the time when the first flow burst actually joins queue $2$ is
complicated by the fact that $x_{2}(t)$ evolves while this burst is in transit. This is illustrated through the example in
Fig. \ref{SFMwithdelay} which we will use to describe the evaluation of this
time through a sequence of events denoted by $\{J_{1},\ldots,J_{K}\}$ with
associated event times $\{\sigma_{1},\ldots,\sigma_{K}\}$. We define $J_{0}$
to be the event when the flow burst leaves node $1$, i.e., the occurrence of a
switch from $G_{1}(t^{-})=0$ to $G_{i}(t)=1$, and let $\sigma_{0}$ be its
associated occurrence time.
Therefore, an estimate of the time when the flow burst joins the tail of queue
$2$ is given by\vspace{0pt} $\sigma_{1}=\sigma_{0}+[L-x_{2}(\sigma
_{0})]/v(\sigma_{0})$ where $v(\sigma_{0})$ is the \textquotedblleft
speed\textquotedblright\ of the flow burst which we assume to be constant and,
for notational simplicity, set it to $v(\sigma_{0})=1$ (it will become clear
in the sequel that this can be relaxed and treated as random in the context of
IPA). Thus, we define $J_{1}$ to be the event at time $\sigma_{1}$ when the
flow burst covers the distance $L-x_{2}(\sigma_{0})$. In general, however,
$x_{2}(\sigma_{1})\leq{x}_{2}(\sigma_{0})\equiv\bar{x}_{2}(\sigma_{1})$, i.e.,
the \emph{estimate} $\bar{x}_{2}(\sigma_{1})$ of $x_{2}(\sigma_{1})$ is based
on the assumption that $x_{2}(t)$ remains unchanged over $(\sigma_{0}%
,\sigma_{1})$. This is illustrated in the example of Fig. \ref{SFMwithdelay},
where $\dot{x}_{2}(t)=-\beta_{2}(t)<0$ for some $t\in(\sigma_{0},\sigma_{1})$.
Thus, unless $x_{2}(\sigma_{1})={x}_{2}(\sigma_{0})$, we repeat at
$t=\sigma_{1}$ the same process of estimating the time of the next opportunity
that the flow burst might join queue $2$ at time $\sigma_{2}$ to cover the
distance $\bar{x}_{2}(\sigma_{1})-x_{2}(\sigma_{1})$ and define this potential
\emph{joining event} as $J_{2}$. \ref{SFMwithdelay}. This
process continues until event $J_{K}$ occurs at time $\sigma_{K}$, the last
event in the sequence $\{J_{1},\ldots,J_{K}\}$ when $\bar{x}_{2}(\sigma
_{K})=x_{2}(\sigma_{K})$. Note that $J_{K}$ may occur either when $(i)$
$\bar{x}_{2}(\sigma_{K})=x_{2}(\sigma_{K})>0$, in which case the estimate
$\bar{x}_{2}(\sigma_{K})$ incurs no error because $x_{2}(\sigma_{K}%
)=x_{2}(\sigma_{K-1})$, i.e., the queue length at node $2$ remained unchanged
because $\beta_{2}(t)=0$ for $t\in\lbrack\sigma_{K-1},\sigma_{K}]$, or $(ii)$
$\bar{x}_{2}(\sigma_{K})=x_{2}(\sigma_{K})=0$, in which case the flow burst
joins node $2$ while this queue is empty. Since in practice the queues and
flow bursts may consist of discrete entities (e.g., vehicles), we define event
$J_{K}$ as occurring when $\bar{x}_{2}(t)-x_{2}(t)\leq\epsilon$ for some
predefined fixed small $\epsilon$, i.e., a flow burst joins the downstream
queue whenever it is sufficiently close to it. 
The following lemma asserts that the event time sequence $\{\sigma_{1}%
,\ldots,\sigma_{K}\}$ is finite.

\textbf{Lemma 1. }Under the assumption that $J_{K}$ is defined through
$\bar{x}_{2}(\sigma_{K})-x_{2}(\sigma_{K})\leq\epsilon$, the number of events
$K$ in $\{J_{1},\ldots,J_{K}\}$ is bounded. Moreover, its event time
$\sigma_{K}$ is also bounded.

\textbf{Proof:} Observe that $x_{2}(t)\leq L$, since the content of queue $2$
is limited by the physical distance $L$. In addition, $\bar{x}_{2}%
(t)-x_{2}(t)>\epsilon$ prior to event $J_{K}$. It follows that $K\leq
L/\epsilon$. Moreover, in the worst case, a flow burst travels the finite
distance $L$ to find $x_{2}(\sigma_{K})=0$, therefore, $\sigma_{K}\leq
\sigma_{0}+L-x_{2}(\sigma_{0})$. $\blacksquare$

We now formalize the dynamics of the flow transit process
described above. First, the dynamics of $\bar{x}_{2}(t)$, the estimated queue
length when an event $J_{k}$\ occurs, are given by
\begin{align}
\dot{\bar{x}}_{2}(t) &  =0\label{x2bar_dynamics}\\
\bar{x}_{2}(t^{+}) &  =x_{2}(t)\text{ \ if }t=\sigma_{k},\text{ }%
k=1,...,K.\vspace{0pt}\nonumber
\end{align}
with $\bar{x}_{2}(\sigma_{1})=L-x_{2}(\sigma_{0})$ and $\sigma_{0}$ defined
above as the occurrence time of a switch from $G_{1}(t^{-})=0$ to $G_{i}%
(t)=1$. The dynamics of $x_{12}(t)$ are given by
\begin{align}
\dot{x}_{12}(t) &  =\left\{
\begin{array}
[c]{l}%
0\\
\alpha_{1}(t)\\
h_{1}(t)
\end{array}
\right.
\begin{array}
[c]{l}%
\text{if }G_{1}(t)=0\\
\text{if }x_{1}(t)=0,\text{ }\alpha_{1}(t)\leq\beta_{1}(t)\\
\text{otherwise}%
\end{array}
\label{x12_dynamics}\\
x_{12}(t^{+}) &  =0\text{ \ \ if }t=\sigma_{K}\nonumber
\end{align}
The dynamics of $x_{2}(t)$ are no longer described by (\ref{x_dynamics}),
since the queue content is only updated when a flow burst joins queue 2 at
time $\sigma_{K}$. Instead, they are given by%
\begin{align}
\dot{x}_{2}(t)  & =\left\{  \vspace*{0pt}%
\begin{array}
[c]{l}%
-\beta_{i}(t)\\
0
\end{array}
\right.
\begin{array}
[c]{l}%
\text{if }x_{2}(t)>0\text{ and }G_{2}(t)=1\\
\text{otherwise}%
\end{array}
\label{x2dynamics}\\
x_{2}(\sigma_{K}^{+})  & =x_{2}(\sigma_{K})+x_{12}(\sigma_{K})\nonumber
\end{align}
Note that in (\ref{x2bar_dynamics}) and (\ref{x12_dynamics}) the values of
event times $\{\sigma_{1},\ldots,\sigma_{K}\}$ remain unspecified. In order to
provide this specification, we define $\delta_{12}(t)=\bar{x}_{2}(t)-x_{2}(t)$
to be the distance between the head of the flow burst and the tail of
$x_{2}(t)$. Then, observe that $\sigma_{k}=\sigma_{k-1}+\tau(\delta
_{12}(\sigma_{k-1}))$, where $\tau(r)$ is the time to complete a distance
$r\in(0,L]$ and $k=1,...,K-1$. Similar to the clock $z_{i}(t)$ in
(\ref{zi_dynamics}) that dictates the timing of the controlled switching
process, we associate a clock $z_{12}(t)$ to the timing of events in
$\{J_{1},\ldots,J_{K}\}$ as follows:%
\begin{align}
\dot{z}_{12}(t) &  =\left\{
\begin{array}
[c]{l}%
1\\
0
\end{array}
\right.
\begin{array}
[c]{l}%
\text{if }{\delta_{12}(t)>0}\\
\text{otherwise}%
\end{array}
\label{z12_dynamics}\\
z_{12}(t^{+}) &  =0\text{ \ \ if }z_{12}(t)=\tau({\delta_{12}(t)})\nonumber
\end{align}
with an initial condition $z_{12}(\sigma_{0})=0$ and
\begin{align}
\dot{\delta}_{12}(t) &  =0\label{delta12_dynamics}\\
\delta_{12}(t^{+}) &  =\left\{
\begin{array}
[c]{l}%
L-x_{2}(t)\\
\bar{x}_{2}(t)-x_{2}(t)
\end{array}
\right.
\begin{array}
[c]{l}%
\text{if }t=\sigma_{0}\\
\text{if }t=\sigma_{k},k=1,...,K.
\end{array}
\vspace*{0pt}\nonumber
\end{align}
Note that $\delta_{12}(t)$ is piecewise constant and updated only at the times
when events $J_{0},J_{1},\ldots,J_{K}$ take place ending with $\delta
_{12}(t^{+})=0$ when event $J_{K}$ occurs, i.e., the flow burst joins queue
$2$. The values of $\tau({\delta_{12}(t)})$ in (\ref{z12_dynamics}) are given
by the time required for the flow burst to travel a distance ${\delta
	_{12}(t)=}\bar{x}_{2}(t)-x_{2}(t)$ with speed $v(\sigma_{0})$ which we assumed
earlier to be constant and set to $v(\sigma_{0})=1$. Thus, $\tau
(\delta_{12}(t))=\delta_{12}(t)$.
Finally, note that in this modeling framework, we assume that $x_{2}(t)$ is
observable at event times $\sigma_{0},\sigma_{1},\ldots,\sigma_{K}$ when
events $J_{0},J_{1},\ldots,J_{K}$ take place.

As a final step, we generalize this model to include multiple flow bursts that
may be generated in an interval $(t_{1},t_{2}]$ such that $G_{1}(t)=1$ for
$t\in\lbrack t_{1},t_{2})$, $G_{1}(t_{1}^{-})=0$. Thus, we denote by
$J_{k}^{n}$ the $k$th event for the $n$th flow burst to (potentially) join
queue $2$ and extend $\delta_{12}(t)$ to $\delta_{12}^{n}(t)$, $\sigma_{k}$ to
$\sigma_{k}^{n}$, and $x_{12}(t)$ to $x_{12}^{n}(t)$, $n=1,2,\ldots$ Also, we
define $J_{i,j}$ as an event such that the $i$th flow burst merges with
the $j$th burst at time $\tau_{i,j}$. For simplicity, we use $y^{m}(t)$ to
represent $x_{12}^{m}(t)$. We then have:\vspace*{-0.5pt}
\begin{align}
\dot{x}_{12}^{n}(t) &  =\left\{
\begin{array}
[c]{l}%
\alpha_{1}(t)\\
h_{1}(t)\\
\\
0\\
\end{array}
\right.
\begin{array}
[c]{l}%
\text{if }n=1,x_{1}(t)=0,\text{ }\alpha_{1}(t)<\beta_{1}(t)\\
\text{if }n=1,G_{1}(t)=1\\
x_{1}(t)=0,\text{ }\alpha_{1}(t)\geq\beta_{1}(t)\text{ or }x_{1}(t)>0\\
\text{otherwise}%
\end{array}
\\
x_{12}^{n}(t^{+}) &  =0\text{ \ \ if }t=\sigma_{K}^{n}\text{ or }%
t=\tau_{n,n-1}\\
x_{12}^{n}(t^{+}) &  =x_{12}^{n}(t)+x_{12}^{n-1}(t)\text{ \ \ if }%
t=\tau_{n+1,n}\nonumber
\end{align}

\begin{align}
\dot{\delta}_{12}^{n}(t) &  =0\label{delta12gen_dynamics}\\
\delta_{12}^{n}(t^{+}) &  =\left\{
\begin{array}
[c]{l}%
L-x_{2}(t)\\
\bar{x}_{2}^{n}(t)-x_{2}(t)\\
\delta_{12}^{n}(t)-y^{m}(t)
\end{array}
\right.
\begin{array}
[c]{l}%
\text{if }t=\sigma_{0}^{n}\\
\text{if }t=\sigma_{k}^{n},k>0\\
\text{if }t=\sigma_{K}^{m},m=1,\ldots,n-1
\end{array}
\nonumber
\end{align}%
\begin{align}
\dot{\bar{x}}_{2}^{n}(t) &  =0\label{x2bargen_dynamics}\\
\bar{x}_{2}^{n}(t^{+}) &  =\left\{
\begin{array}
[c]{l}%
x_{2}(t)\\
x_{2}(t)+y^{m}(t)
\end{array}
\right.
\begin{array}
[c]{l}%
\text{if }t=\sigma_{k}^{n},k\geq0\\
\text{if }t=\sigma_{K}^{m},m=1,\ldots,n-1
\end{array}
\nonumber
\end{align}
with the obvious generalizations of (\ref{x2bar_dynamics})-(\ref{delta12_dynamics}).
\begin{figure}[pt]
	\centering
	\includegraphics[scale=0.4]{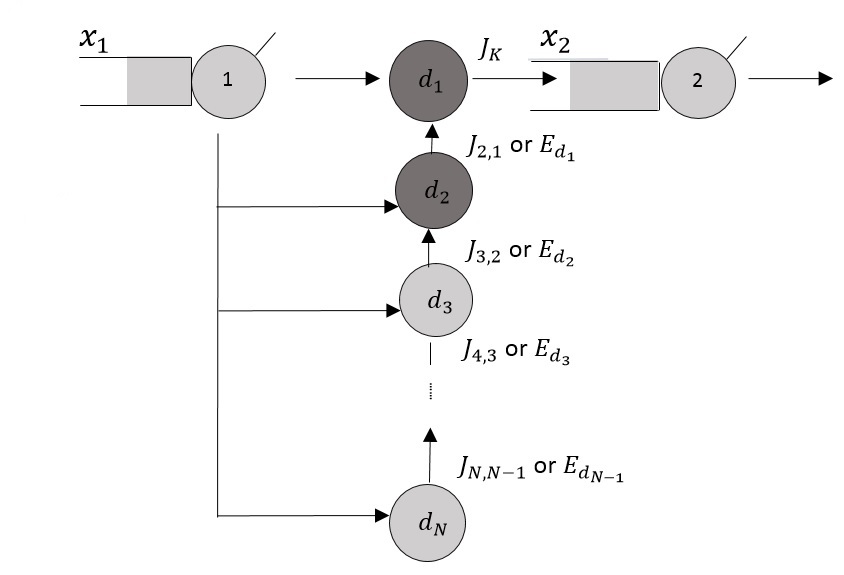} \caption{Two-node SFM with delay.}%
	\label{twonodeSFMdelay}%
\end{figure}
The generalized SFM with delay is shown in Fig.
\ref{twonodeSFMdelay}. We define a series of servers $d_{n}$, $n\in
\{j\in\mathbb{Z}:j=1,\ldots,N\}$ to describe the flow transit delay between
SFM where $y_{n}(t)$ is the content of $d_{n}$. Here, $N$ is the total number
of servers required depending on a specific application. For example, in the
two-intersection traffic system discussed in the next section, we set
$N=\lceil L/L_{v}\rceil$ where $L$ is the physical distance between
intersections and $L_{v}$ is the length of a vehicle. When a new flow burst
leaves server $1$, the controlled switching process checks whether
$y_{1}(t)=0$ to initiate a flow burst. If $y_{1}(t)>0$, it checks $y_{j}(t)$
for $j\geq2$ until some $y_{j}(t)=0$. 
For example, in Fig.
\ref{twonodeSFMdelay}, if servers $d_{1}$ and $d_{2}$ are non-empty (dark
color), and $d_{3}$ is empty (light color), the new flow burst will join
server $d_{3}$ until $y_{1}(t)=0$. The first flow burst will leave server
$d_{1}$ when event $J_{K}^{1}$ occurs and joins $x_{2}(t)$. The flow burst in
server $d_{n}$ will leave when either one of two events occurs, defined as
follows: (1) $J_{n,n-1}$ occurs when the $n$th flow burst joins the $(n-1)$th
burst. (2) $E_{d_{n-1}}$ occurs when $y_{n-1}(t)=0$.

\textbf{SFM Events}. The hybrid system with dynamics given by
(\ref{x_dynamics})-(\ref{delta12_dynamics}) defines the SFM with transit
delays. To complete the model, we define next the event set associated with
all discontinuous state transitions in (\ref{x_dynamics}%
)-(\ref{delta12_dynamics}). As in prior work using SFMs, we observe that the
sample path of any queue content process in our model can be partitioned into
\emph{Non-Empty Periods} (NEPs) when $x_{i}(t)>0$, and \emph{Empty Periods}
(EPs) when $x_{i}(t)=0$. Let us define the start of a NEP at queue $i$ as
event $S_{i}$ ($S_{12}$ for queue $12$) and the end of a NEP at queue $i$ as
event $E_{i}$ ($E_{12}$ for queue $12$). In (\ref{x_dynamics}), observe that
$S_{1}$ is an event that can be induced by either an event such that
$\alpha_{1}(t)-\beta_{2}(t)$ switches from $\leq0$ to $>0$ or by an event
which switches the value of $\beta_{1}(t)$; moreover, in
(\ref{SFMbeta_dynamics}), the value of $\beta_{1}(t)$ switches when an event
occurs such that $G_{1}(t)$ changes between $0$ and $1$. In
(\ref{x2dynamics}), $S_{2}$ may also be induced by event $J_{k}$ if it occurs
when $x_{2}(t)=0$. Finally, in (\ref{x12_dynamics}), $S_{12}$ is induced by
the same events that induce $S_{2}$, while $E_{12}$ is induced by $J_{K}$
since that causes the end of the flow burst that created $x_{12}(t)>0$. To sum
up, there are five events that can affect any of the processes $\{x_{1}(t)\}$,
$\{x_{2}(t)\}$ and $\{x_{12}(t)\}$:

1. $E_{i}$: $x_{i}(t)$ switches from $>0$ to $=0$, thus ending a NEP at queue
$i$.

2. $\Gamma_{i}$: $\alpha_{i}(t)-\beta_{i}(t)$ switches from $\leq0$ to $>0$. 

3. $J_{k}$: $z_{12}(t)=\tau({\delta_{12}(t)})$ representing a potential
joining of the flow burst $x_{12}(t)$ with $x_{2}(t)$ if ${\delta_{12}
	(t^{+})>0}$, or the actual joining if ${\delta_{12}(t^{+})=0}$.

4. $C2O_{i}$: $G_{i}(t)$ switches from $1$ to $0$.

5. $O2C_{i}$: $G_{i}(t)$ switches from $0$ to $1$.

We can now identify the event set that affects the dynamics of the three
queue content processes:\vspace{0pt}
\begin{align*}
\Phi_{1} &  =\{S_{i},E_{i},O2C_{i},C2O_{i}\}\text{,}\\
\Phi_{2} &  =\{S_{2},E_{2},O2C_{2},C2O_{2},J_{k}\},\text{ }\Phi_{12}%
=\{S_{12},E_{12},E_{1},C2O_{1},J_{k}\}
\end{align*}
Finally, note that this SFM model can be extended to any network of queues
with possible delays by identifying queues with dynamics of type
(\ref{x_dynamics}) or (\ref{x2dynamics}) or (\ref{x12_dynamics}).

\section{MULTI-INTERSECTION TRAFFIC LIGHT CONTROL WITH DELAYS \label{3}}

\begin{figure}[pt]
\centering
\includegraphics[scale=0.4]{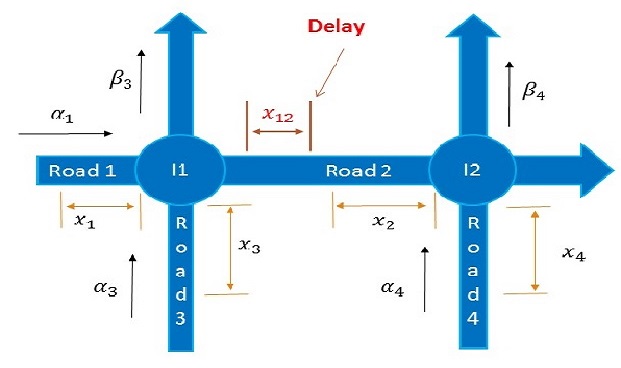} \caption{Two traffic
intersections.}%
\label{Twotrafficintersections}%
\end{figure}

An application of the SFM with delays arises in the Traffic Light Control
(TLC) problem in transportation networks, which consists of adjusting green
and red signal settings in order to control the traffic flow through an
intersection and, more generally, through a set of intersections and traffic
lights in an urban roadway network. The ultimate objective is to minimize
congestion in an area consisting of multiple intersections. Many methods have
been proposed to solve the TLC problem, including expert systems, genetic
algorithms, reinforcement learning and several optimization techniques; a more
detailed review of such methods may be found in \cite{fleck2016adaptive}.
Perturbation analysis methods were used in \cite{Head1996} and \cite{Fu2003}.
IPA was used in \cite{Panayiotou2005} and \cite{geng2012traffic} for a single
intersection and extended to multiple intersections in \cite{geng2015multi}
and to quasi-dynamic control schemes in \cite{fleck2016adaptive}. However, all
this work to date has assumed that vehicles moving from one intersection to
the next experience no delay. In this section, we formulate the TLC problem by
including delays as in Section 2 and derive an IPA-based controller to
optimize selected performance metrics (cost functions). By including delays,
we will see that we can define new metrics which capture \textquotedblleft
congestion\textquotedblright\ in traffic systems much more accurately.
\begin{figure}[ptb]
\begin{center}
\includegraphics[scale=0.27]{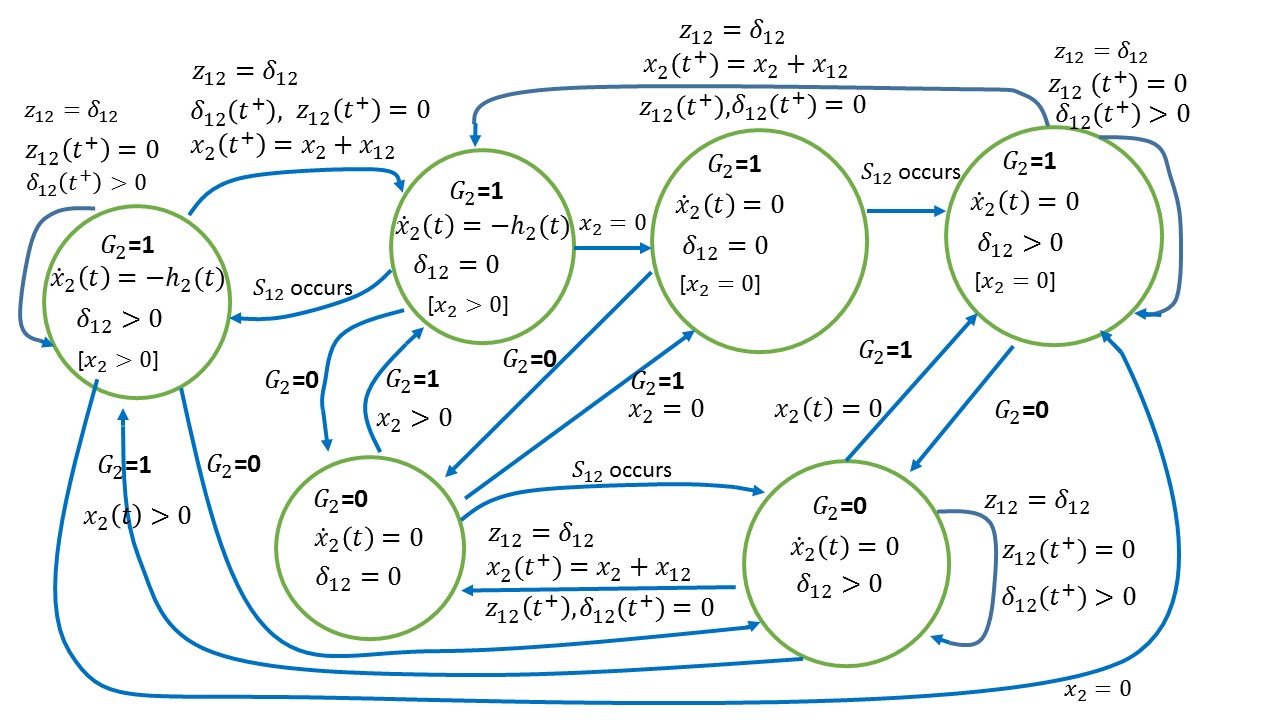}
\end{center}
\caption{Stochastic Hybrid Automaton model for $x_{2}(t)$.}%
\label{sha_x2}%
\end{figure}
As in Section 2, let $\{\alpha_{i}(t)\}$ and $\{\beta_{i}(t)\}$,
$i=1,\ldots,4$, be the incoming and outgoing flow processes\emph{
}respectively at all four roads shown in Fig. \ref{Twotrafficintersections},
where we now interpret $\alpha_{i}(t)$ as the random instantaneous vehicle
arrival rate at time $t$. We define the controllable parameters $\theta_{i}$
to be the durations of the GREEN light for road $i=1,\ldots,4$. Thus, the
state vector is $x(\theta,t)=[x_{1}(\theta,t),x_{2}(\theta,t),x_{3}%
(\theta,t),x_{4}(\theta,t),x_{12}(\theta,t)]$ where $x_{i}(\theta,t)$ is the
content of queue $i$ and $x_{12}(\theta,t)$ is the content of the road between
intersections $I1$ and $I2$. To maintain notational simplicity, we will assume
in our analysis that (\textbf{A1}) There is no more than one traffic burst in
queue $12$ at any one time, (\textbf{A2}) The speed of a traffic burst
$v_{1}(t)$ between intersections is constant, and (\textbf{A3}) There is no
traffic coupling between $I1$ and $I2$. Assumptions (\textbf{A1}) and
(\textbf{A2}) simplify the analysis and can be easily relaxed since our model
can deal with multiple flow bursts as shown in Section 2. Assumption
(\textbf{A3}) means that the distance between $I2$ and $I1$ is sufficiently
large and is also made to simplify the model; it can be relaxed along the
lines of \cite{geng2015multi}.

We define clock state variables $z_{i}(t)$, $i=1,\ldots,4$, which are
associated with the GREEN light cycle for queue $i$ based on
(\ref{zi_dynamics}) where the controller $G_{i}(t)$ is now the traffic light
state, i.e., $G_{i}(t)=0$ means that the traffic light in road $i$ is RED,
otherwise, it is GREEN. Accordingly, the departure rates and the queue content
dynamics $x_{i}(t)$, $i=1,\ldots,4$, are given by 
(\ref{x_dynamics})-(\ref{x2dynamics}).

In order to provide the dynamics of $x_{2}(t)$ and $x_{12}(t)$, we will make
use of our analysis in Section 2. In particular, let $\sigma_{0}$ be the time
when a positive traffic flow is generated from queue $1$ and enters queue
$12$, i.e., the light turns from RED to GREEN for road $1$ and $x_{1}%
(\sigma_{0})>0$. Invoking (\ref{delta12_dynamics}), we define $\delta_{12}(t)$
to be the distance between the head of the \textquotedblleft transit
queue\textquotedblright\ $12$ and the tail of queue $2$. Thus, $\delta
_{12}(\sigma_{0}^{+})=L-x_{2}(\sigma_{0})$. We also associate a clock to this
queue, denoted by $z_{12}(t)$, which is defined by (\ref{z12_dynamics}) and
initialized at $z_{12}(\sigma_{0})=0$. Finally, $\tau(\delta_{12}(t))$ in
(\ref{z12_dynamics}) in the TLC context is given by $\tau(\delta
_{12}(t))=\delta_{12}(t)/v_{1}$.

Recall that a $J_{k}$ event represents a \emph{potential} joining of the flow
burst from $I1$ with queue $2$. The \emph{actual} joining event occurs when
${\delta_{12}(t^{+})=0}$ from its initial value $\delta_{12}(\sigma_{0}%
^{+})=L-x_{2}(\sigma_{0})$. Adapting (\ref{delta12_dynamics}) and
(\ref{x2bar_dynamics}) to the TLC setting we get the dynamics of $\delta_{12}$
and $\bar{x}_{2}(t)$, while the dynamics of $x_{2}(t)$ and $x_{12}(t)$ are
given by (\ref{x2dynamics}) and (\ref{x12_dynamics}) respectively.

\textbf{SFM Events}. We apply the event set defined in Section 2 where we use
$G2R_{i}$ (traffic light $i$ changes from GREEN to RED) to replace $C2O_{i}$
and $R2G_{i}$ to replace $O2C_{i}$. Figure \ref{sha_x2} shows the hybrid
automaton model for queue 2 in terms of its six possible modes depending on
$x_{2}(t)$, $G_{2}(t)$ and $\delta_{12}(t)$. Similar models apply to the
remaining processes, all of which are generally interdependent(e.g., in Fig.\ref{sha_x2}, some reset conditions involve $x_{12}(t)$).

\textbf{Cost Functions}. The objective of the TLC problem is to control the
green cycle parameters $\theta_{i}$, $i=1,\ldots,4$, so as to minimize traffic
congestion in the region covered by the two intersections in Fig.
\ref{Twotrafficintersections}. In \cite{geng2012traffic} and
\cite{fleck2016adaptive}, the average total weighted queue lengths over a
fixed time interval $[0,T]$ is used to capture congestion:\vspace{0pt}
\begin{equation}
F(\theta;x(0),z(0),T)=\frac{1}{T}\sum_{i=1}^{5}\int_{0}^{T}w_{i}x_{i}%
(\theta,t)dt. \label{ave_cost}%
\end{equation}
where $w_{i}$ is the weight associated with queue $i$. For convenience, we
will refer to \eqref{ave_cost} as the \emph{average queue} cost function; with
a slight abuse of notation we have re-indexed $x_{12}(t)$ as $x_{5}(t)$.
However, this may not be an adequate measure of \textquotedblleft
congestion\textquotedblright. For instance, it is possible that the average
queue lengths over $[0,T]$ are relatively small, while reaching large values
over small intervals (peak periods during a typical day). Thus, instead of
restricting ourselves to \eqref{ave_cost}, we define next two new cost functions.

\textbf{1.} \emph{Average weighted }$P$\emph{th power of the queue lengths}
over a fixed interval $[0,T)$, where $P>1$. The sample function is\vspace
{-\belowdisplayskip}
\[
F(\theta;x(0),z(0),T)=\frac{1}{T}\sum_{i=1}^{5}\int_{0}^{T}w_{i}x_{i}%
^{P}(\theta,t)dt.
\]
Observing that $x_{i}(\theta,t)=0$ during an EP of queue $i$, we can rewrite
this as\vspace{-\belowdisplayskip}
\begin{equation}
F(\theta;x(0),z(0),T)=\frac{1}{T}\sum_{i=1}^{5}\sum_{m=1}^{M_{i}}\int
_{\xi_{i,m}}^{\eta_{i,m}}w_{i}x_{i}^{P}(\theta,t)dt, \label{power_cost1}%
\end{equation}
in which $M_{i}$ is the total number of NEPs of queue $i$ over a time interval
$[0,T]$ and $\xi_{i,m}$, $\eta_{i,m}$ are the occurrence times of the $m$th
$S_{i}$ event and $E_{i}$ event respectively. We also define the cost incurred
within the $m$th NEP of queue $i$ as\vspace*{-\belowdisplayskip}
\begin{equation}
F_{i,m}(\theta)=\int_{\xi_{i,m}}^{\eta_{i,m}}w_{i}x_{i}^{P}(\theta,t)dt.
\label{power_cost2}%
\end{equation}
Clearly, when $P=1$, \eqref{power_cost1} is reduced to \eqref{ave_cost}. When
$P>1$, \eqref{power_cost1} amplifies the presence of intervals where queue
lengths are large. Therefore, minimizing \eqref{power_cost1} decreases the
probability that a road develops a large queue length. We will refer to this
metric \eqref{power_cost1} as the \emph{power cost function.}

\textbf{2.} \emph{Average} \emph{weighted fraction of time that queue lengths
exceed given thresholds} over a fixed interval $[0,T]$. The sample function is
\vspace{-\belowdisplayskip}
\begin{align}
F(\theta;x(0),z(0),T)  &  =\frac{1}{T}\sum_{i=1}^{5}\int_{0}^{T}%
w_{i}\mathbf{1}[x_{i}(\theta,t)>\zeta_{i}]dt\label{th_cost1}\\
&  =\frac{1}{T}\sum_{i=1}^{5}\int_{0}^{T}w_{i}r_{i}(\theta,t)dt \vspace
*{-\baselineskip}\nonumber
\end{align}
where $\zeta_{i}$ is a given threshold and $r_{i}(\theta,t)=\mathbf{1}%
[x_{i}(\theta,t)>\zeta_{i}]$. This necessitates the definition of two
additional events: $Z_{i}$ is the event such that $x_{i}(\theta,t)=\zeta_{i}$,
$x_{i}(\theta,t^{-})<\zeta_{i}$ (i.e., the queue content reaches the threshold
from below) and $\bar{Z}_{i}$ is the event such that $x_{i}(\theta
,t)<\zeta_{i}$, $x_{i}(\theta,t^{-})=\zeta_{i}$. Observe that $\dot{r}%
_{i}(\theta,t)=0$ with a reset condition $r_{i}(\theta,t^{+})=1$ if
$x_{i}(\theta,t^{-})<\zeta_{i}$, $x_{i}(\theta,t^{+})=\zeta_{i}$ and
$r_{i}(\theta,t^{+})=0$ if $x_{i}(\theta,t^{-})=\zeta_{i}$, $x_{i}%
(\theta,t^{+})<\zeta_{i}.$ Finally, we use $F_{i,m}(\theta)$ as in
(\ref{power_cost2}), for the cost associated with the $m$th NEP at queue
$i$:\vspace{-\belowdisplayskip}%
\begin{equation}
F_{i,m}(\theta)=\int_{\gamma_{i,m}(\theta)}^{\psi_{i,m}(\theta)}w_{i}%
r_{i}(\theta,t)dt. \vspace{-0.6pt} \label{th_cost2}%
\end{equation}
where $\gamma_{i,m}$, $\psi_{i,m}$ are the start and end respectively of an
interval such that $r_{i}(\theta,t)=1$.

\textbf{Optimization}. Our purpose is to minimize the cost functions defined
in \eqref{ave_cost}, \eqref{power_cost1} and \eqref{th_cost1}. We define the
overall cost function as follows:\vspace{-0.6pt}
\[
H(\theta;x(0),z(0),T)=E[F(\theta;x(0),z(0),T)],
\]
in which $F(\theta;x(0),z(0),T)$ is a sample cost function of the form
\eqref{ave_cost}, \eqref{power_cost1} or \eqref{th_cost1}. Clearly, we cannot
derive a closed-form expression for the expectation above. However, we can
estimate the gradient $\nabla H(\theta)$ through the sample gradient $\nabla
F(\theta)$ based on IPA, which has been shown to be unbiased under mild
technical conditions (Proposition 1 in \cite{Cassandras2010}). We emphasize
that \emph{no explicit knowledge of }$\alpha_{i}(t)$\emph{ and }$h_{i}%
(t)$\emph{ is necessary} to estimate $\nabla H(\theta)$. The IPA estimators
derived in the next section only need estimates of $\alpha_{i}(\tau_{k})$ and
$h_{i}(\tau_{k})$ at certain event times $\tau_{k}$. Using $\nabla F(\theta)$,
we can use a simple gradient-descent optimization algorithm to minimize the
associated cost metric through the iterative scheme\vspace*{-0.7pt}
\[
\theta_{j,k+1}=\theta_{j,k}-c_{k}Q_{j,k}(\theta_{k},x(0),T,\omega_{k}),
\]
in which $Q_{j,k}(\theta_{k},x(0),T,\omega_{k})$ is an estimator of
$dH/d\theta_{j}$ (in our case, $dF/d\theta_{j}$) in sample path $\omega_{k}$
and $c_{k}$ is the step size at the $k$th iteration selected through an
appropriate decreasing sequence to guarantee convergence
(\cite{fleck2016adaptive}). In the next section, we use the IPA methodology to
obtain $dF/d\theta_{j}$ through the state derivatives $\frac{\partial
x_{i}(\theta,t)}{\partial\theta_{j}}$.

\section{INFINITESIMAL PERTURBATION ANALYSIS (IPA)}

We briefly review the IPA framework for general stochastic hybrid systems as
presented in \cite{Cassandras2010}. Let $\{\tau_{k}(\theta)\}$, $k=1,\ldots
,K$, denote the occurrence times of all events in the state trajectory of a
hybrid system with dynamics $\dot{x}\ =\ f_{k}(x,\theta,t)$ over an interval
$[\tau_{k}(\theta),\tau_{k+1}(\theta))$, where $\theta\in\Theta$ is some
parameter vector and $\Theta$ is a given compact, convex set. For convenience,
we set $\tau_{0}=0$ and $\tau_{K+1}=T$. We use the Jacobian matrix notation:
$x^{\prime}(t)\equiv\frac{\partial x(\theta,t)}{\partial\theta}$ and $\tau
_{k}^{\prime}\equiv\frac{\partial\tau_{k}(\theta)}{\partial\theta}$, for all
state and event time derivatives. It is shown in \cite{Cassandras2010} that
\begin{equation}
\frac{d}{dt}x^{\prime}(t)=\frac{\partial f_{k}(t)}{\partial x}x^{\prime
}(t)+\frac{\partial f_{k}(t)}{\partial\theta}, \label{eq:IPA_1}%
\end{equation}
for $t\in\lbrack\tau_{k},\tau_{k+1})$ with boundary condition:
\begin{equation}
x^{\prime}(\tau_{k}^{+})=x^{\prime}(\tau_{k}^{-})+[f_{k-1}(\tau_{k}^{-}%
)-f_{k}(\tau_{k}^{+})]\tau_{k}^{\prime} \label{eq:IPA_2}%
\end{equation}
for $k=1,...,K$. In order to complete the evaluation of $x^{\prime}(\tau
_{k}^{+})$ in (\ref{eq:IPA_2}), we need to determine $\tau_{k}^{\prime}$. If
the event at $\tau_{k}$ is \emph{exogenous} (i.e., independent of $\theta$),
$\tau_{k}^{\prime}=0$. However, if the event is \emph{endogenous}, there
exists a continuously differentiable function $g_{k}:\mathbb{R}^{n}%
\times\Theta\rightarrow\mathbb{R}$ such that $\tau_{k}\ =\ \min\{t>\tau
_{k-1}\ :\ g_{k}\left(  x\left(  \theta,t\right)  ,\theta\right)  =0\}$ and,
as long as $\frac{\partial g_{k}}{\partial x}f_{k}(\tau_{k}^{-})\neq0$,
\begin{equation}
\tau_{k}^{\prime}=-\left[  \frac{\partial g_{k}}{\partial x}f_{k}(\tau_{k}%
^{-})\right]  ^{-1}\left[  \frac{\partial g_{k}}{\partial\theta}%
+\frac{\partial g_{k}}{\partial x}x^{\prime}(\tau_{k}^{-})\right]
\label{eq:IPA_3}%
\end{equation}
In our TLC setting, we will use the notation
\[
{x}_{i,j}^{^{\prime}}(t)=\frac{\partial x_{i}(\theta,t)}{\partial\theta_{j}%
},{z}_{i,j}^{^{\prime}}(t)=\frac{\partial z_{i}(\theta,t)}{\partial\theta_{j}%
},{\tau}_{k,j}^{^{\prime}}(t)=\frac{\partial\tau_{k}(\theta)}{\partial
\theta_{j}}%
\]
We also note that in (\ref{x_dynamics}),(\ref{x12_dynamics}), $\frac{\partial
f_{k}(t)}{\partial\theta}=\frac{\partial f_{k}(t)}{\partial x}=0$ and
(\ref{eq:IPA_1}) reduces to\vspace{-1pt}%
\begin{equation}
x_{i,j}^{^{\prime}}(t)=x_{i,j}^{^{\prime}}(\tau_{k}^{+}),\text{ \ \ \ }%
t\in(\tau_{k},\tau_{k+1}] \label{eq:IPA_1reduced}%
\end{equation}

\subsection{State and Event Time Derivatives}

We will now apply the IPA equations (\ref{eq:IPA_2})-(\ref{eq:IPA_1reduced})
to our TLC setting on an event by event basis for each of the events sets
$\Phi_{i}$, $i=1,\ldots,4$, and $\Phi_{12}$. In all cases, $\tau_{k}$ denotes
the associated event time.

\subsubsection{\textit{IPA for Event Set }$\Phi_{i}=\{S_{i},E_{i}
,R2G_{i},G2R_{i}\}\cup\{Z_{i}$,$\bar{Z}_{i}\}$, $i=1,3,4$}

IPA for these three processes for each of the events in the first set above is
identical to that in \cite{geng2012traffic}. Thus, we simply summarize the
results here.

\textbf{(1)} \textit{Event }$E_{i}$: $x_{i,j}^{^{\prime}}(\tau_{k}^{+})=0$.

\textbf{(2)} \textit{Event }$G2R_{i}$:\textit{ }Let $\rho_{k}$ be the time of
the last $R2G_{i}$ event before $G2R_{i}$ occurs. Then, $\tau_{k,j}^{^{\prime
}}=\mathbf{1}[j=i]+\rho_{k,j}^{^{\prime}}$ and\vspace{-\belowdisplayskip}
\begin{equation}
{x}_{i,j}^{^{\prime}}(\tau_{k}^{+})=\left\{
\begin{array}
[c]{l}%
{x}_{i,j}^{^{\prime}}(\tau_{k})-\alpha_{i}(\tau_{k})\tau_{k,j}^{^{\prime}}\\
{x}_{i,j}^{^{\prime}}(\tau_{k})-h_{i}(\tau_{k})\tau_{k,j}^{^{\prime}}%
\end{array}
\right.
\begin{array}
[c]{l}%
\text{if }x_{i}(t)=0,\text{ }\alpha_{i}(t)\leq\beta_{i}(t)\\
\text{otherwise}%
\end{array}
\label{state_derivative_G2R_i}%
\end{equation}

\textbf{(3)} \textit{Event }$R2G_{i}$:\textit{ }Let $\rho_{k}$ be the time of
this event and $\tau_{k}$ be the time of the last $G2R_{i}$ event before
$R2G_{i}$ occurs. We will use the

notation $\bar{\imath}$ to denote the index of a road perpendicular to $i$
(e.g., $\bar{1}=3$, $\bar{2}=4$). Then, $\rho_{k,j}^{^{\prime}}=\mathbf{1}%
[j=\bar{\imath}]+\tau_{k,j}^{^{\prime}}$ and\vspace{0pt}
\begin{equation}
{x}_{i,j}^{^{\prime}}(\rho_{k}^{+})=\left\{
\begin{array}
[c]{l}%
{x}_{i,j}^{^{\prime}}(\rho_{k})+\alpha_{i}(\rho_{k})\rho_{k,j}^{^{\prime}}\\
{x}_{i,j}^{^{\prime}}(\rho_{k})+h_{i}(\rho_{k})\rho_{k,j}^{^{\prime}}%
\end{array}
\right.
\begin{array}
[c]{l}%
\text{if }x_{i}(t)=0,\text{ }\alpha_{i}(t)\leq\beta_{i}(t)\\
\text{otherwise}%
\end{array}
\label{state_derivative_R2G_i}%
\end{equation}

\textbf{(4)} \textit{Event }$S_{i}$: If $S_{i}$ is induced by $G2R_{i}$, then
${x}_{i,j}^{^{\prime}}(\tau_{k}^{+})={x}_{i,j}^{^{\prime}}(\tau_{k}
)-\alpha_{i}(\tau_{k})\tau_{k,j}^{^{\prime}}$. If $S_{i}$ is an exogenous
event triggered by $\Gamma_{i}$ , then ${x}_{12,j}^{^{\prime}}(\tau_{k}
^{+})={x}_{12,j}^{^{\prime}}(\tau_{k})$.\newline

For the two new events $\{Z_{i}$,$\bar{Z}_{i}\}$, we have:

\textbf{(5)} \textit{Event }$Z_{i}$: This is an endogenous event which occurs
when $g_{k}(x(\theta,t),\theta)=x_{i}(\tau_{k})-\zeta_{i}=0$. Applying
(\ref{eq:IPA_3}), we have\vspace{-\belowdisplayskip}
\begin{equation}
{\tau}_{k,j}^{^{\prime}}=\left\{
\begin{array}
[c]{l}%
-{x}_{i,j}^{^{\prime}}(\tau_{k})/\alpha_{i}(\tau_{k})\\
-{x}_{i,j}^{^{\prime}}(\tau_{k})/[\alpha_{i}(\tau_{k})-h_{i}(\tau_{k})]
\end{array}
\right.
\begin{array}
[c]{l}%
\text{if }G_{i}(t)=0\\
\text{if }G_{i}(t)=1
\end{array}
\label{eventtime_derivative_B_i}%
\end{equation}
Moreover, based on the definition $r_{i}(t)=\mathbf{1}[x_{i}(t)>\zeta_{i}]$ in
Section \ref{3}, $r_{i}(\tau_{k}^{+})=1$, which implies that
$r_{i,j}^{^{\prime}}(\tau_{k}^{+})+\dot{r}(\tau_{k}^{+})\tau_{k,j}^{+}=0$.
Since $\dot{r}_{i}(\tau_{k}^{+})=0$, we get $r_{i,j}^{^{\prime}}(\tau_{k}
^{+})=0$.

\textbf{(6)} \textit{Event }$\bar{Z}_{i}$: Similar to the previous case,
$g_{k}(x(\theta,t),\theta)=x_{i}(\tau_{k})-\zeta_{i}=0$ and applying
(\ref{eq:IPA_3}) gives
\begin{equation}
{\tau}_{k,j}^{^{\prime}}=-{x}_{i,j}^{^{\prime}}(\tau_{k})/(\alpha_{i}(\tau
_{k})-h_{i}((\tau_{k}))) \label{eventtime_derivative_EB_i}%
\end{equation}
In this case, $r_{i}(\tau_{k}^{+})\equiv0$, therefore, $r_{i,j}^{^{\prime}
}(\tau_{k}^{+})+\dot{r}_{i}(\tau_{k}^{+})\tau_{k,j}^{+}=0$ and, since $\dot
{r}_{i}(\tau_{k}^{+})=0$, we get $r_{i,j}^{^{\prime}}(\tau_{k}^{+})=0$.

\subsubsection{\textit{IPA for Event Set }$\Phi_{2}=\{S_{2},E_{2}
,R2G_{2},G2R_{2},J_{k}\}\cup\{Z_{2}$,$\bar{Z}_{2}\}$}

IPA for this set and for $\Phi_{12}$ is different as detailed next.

\textbf{(1)} \textit{Event }$E_{2}$: This is an endogenous event ending an EP
that occurs when $g_{k}(x(\theta,t),\theta)=x_{2}(t)=0$ at $t=\tau_{k}$.
Applying (\ref{eq:IPA_3}) and using (\ref{x2dynamics}), we have $\tau
_{k,j}^{^{\prime}}={x_{2,j}^{^{\prime}}(\tau_{k}^{-})}/{h_{2}(\tau_{k}^{-})}$.
It then follows from (\ref{eq:IPA_2}) that $x_{2,j}^{^{\prime}}(\tau_{k}
^{+})={x_{2,j}^{^{\prime}}(\tau_{k}^{-})-h_{2}(\tau_{k}^{-})}\tau
_{k,j}^{^{\prime}}=0$.

\textbf{(2)} \textit{Event }$S_{2}$: In view of the reset condition in
(\ref{x2dynamics}), this event is induced by $J_{k}$ provided $\delta
_{12}(t^{+})=0$. As described in Section 2, a sequence of $J_{k}$ events is
initiated when a flow burst is generated at node $1$ with associated event
times $\{\sigma_{0},\sigma_{1},\ldots,\sigma_{K}\}$. Event $S_{2}$ is induced
by the last occurrence of a $J_{k}$ event at time $\sigma_{K}$. Thus, our goal
here is to evaluate the IPA derivative ${x}_{2,j}^{^{\prime}}(\sigma_{K}^{+}%
)$. At first sight, it would appear that this requires the complete sequence
$\{{x}_{2,j}^{^{\prime}}(\sigma_{0}^{+}),\ldots,{x}_{2,j}^{^{\prime} }%
(\sigma_{K-1}^{+})\}$ along with event time derivatives $\{\sigma
_{0,j}^{^{\prime}},\ldots,\sigma_{K-1,j}^{^{\prime}}\}$ from which ${x}
_{2,j}^{^{\prime}}(\sigma_{K}^{+})$ can be inferred. However, the following
lemma shows that the only information needed from the full sequence of $J_{k}$
events is $\sigma_{0}^{^{\prime}}$.

\textbf{Lemma 2.} Let $\sigma_{k}$, $k=0,1,\ldots,K$ be the occurrence time of
event $J_{k}$ for a flow burst initiated at $\sigma_{0}$. Then,
\[
\sigma_{k,j}^{^{\prime}}=\frac{-1}{{v_{1}}}[{x_{2,j}^{^{\prime}}(\sigma
_{k-1})+\dot{x}_{2}(\sigma{_{k-1}})\sigma_{k-1,j}^{^{\prime}}}]+\sigma
_{0,j}^{^{\prime}}
\]

\textbf{Proof:} Event $J_{k}$ at $t=\sigma_{k}$ is endogenous and occurs when
$g_{k}(x(\theta,\sigma_{k}),\theta)=z_{12}(\sigma_{k})-\delta_{12}(\sigma
_{k})/v_{1}=0$. Applying (\ref{eq:IPA_3}) and using (\ref{z12_dynamics}%
),(\ref{delta12_dynamics}), we get $\sigma_{k,j}^{^{\prime}}=\delta
_{12,j}^{^{\prime}}(\sigma_{k})/v_{1}-z_{12,j} ^{^{\prime}}(\sigma_{k})$.
Using (\ref{eq:IPA_1reduced}), we have $\delta_{12,j}^{^{\prime}}(\sigma
_{k})=\delta_{12,j}^{^{\prime}}(\sigma_{k-1}^{+})$ and it follows
that\vspace{-\belowdisplayskip}
\begin{equation}
\sigma_{k,j}^{^{\prime}}=\delta_{12,j}^{^{\prime}}(\sigma_{k-1}^{+}
)/v_{1}-z_{12,j}^{^{\prime}}(\sigma_{k}) \label{proof_sigma1}%
\end{equation}
Again applying (\ref{eq:IPA_1reduced}) gives $z_{12,j}^{^{\prime}}(\sigma
_{k})=z_{12,j}^{^{\prime}}(\sigma_{k-1}^{+})$. From (\ref{eq:IPA_2}), in view
of (\ref{z12_dynamics}), we get, for $k=1$, $z_{12,j}^{^{\prime}}(\sigma
_{0}^{+})=-\sigma_{0,j}^{^{\prime}}$. The reset condition in
(\ref{delta12_dynamics}) implies that $\delta_{12}(\sigma_{0}^{+}
)=L-x_{2}(\sigma_{0})$, hence $\delta_{12,j}^{^{\prime}}(\sigma_{0}
^{+})=-{x_{2,j}^{^{\prime}}(\sigma_{0})-\dot{x}_{2}(\sigma_{0})\sigma
	_{0,j}^{^{\prime}}}$. Thus, in this case, (\ref{proof_sigma1}) gives:\vspace
{-\belowdisplayskip}
\begin{equation}
\sigma_{1,j}^{^{\prime}}=\frac{-1}{{v_{1}}}[{x_{2,j}^{^{\prime}}(\sigma
	_{0})+\dot{x}_{2}(\sigma_{0})\sigma_{0,j}^{^{\prime}}}]+\sigma_{0,j}
^{^{\prime}} \label{sigmaprime_k=0}%
\end{equation}
For $k>1$, based on the reset condition in (\ref{z12_dynamics}), we have
$z_{12}(\sigma_{k}^{+})=0$. Taking the total derivative, we get $z_{12,j}
^{^{\prime}}(\sigma_{k}^{+})=-\sigma_{k,j}^{^{\prime}}$. The reset condition
in (\ref{delta12_dynamics}) now implies that $\delta_{12}(\sigma_{k-1}
^{+})=\bar{x}_{2}({\sigma_{k-1}})-x_{2}({\sigma_{k-1}})$, hence\vspace
{-\belowdisplayskip}
\begin{align}
\delta_{12,j}^{^{\prime}}(\sigma_{k-1}^{+})  &  =\bar{x}_{2,j}^{^{\prime}
}(\sigma_{k-1})+\dot{\bar{x}}_{2}(\sigma_{k-1})\sigma_{k-1,j}^{^{\prime}%
}\label{proof_delta12prime}\\
&  -x_{2,j}^{^{\prime}}(\sigma_{k-1})-\dot{x}_{2}(\sigma_{k-1})\sigma
_{k-1,j}^{^{\prime}}\nonumber
\end{align}
Applying (\ref{eq:IPA_1reduced}), we have $\bar{x}_{2,j}^{^{\prime}}%
(\sigma_{k-1})=\bar{x}_{2,j}^{^{\prime}}(\sigma_{k-2}^{+})$. Looking at
(\ref{x2bar_dynamics}), we have $\dot{\bar{x}}_{2}(\sigma_{k-1})=0$ and the
reset condition implies that $\bar{x}_{2,j}^{^{\prime}}(\sigma_{k-2}%
^{+})=x_{2,j}^{^{\prime}}(\sigma_{k-2})+\dot{x}_{2}(\sigma_{k-2}%
)\sigma_{k-2,j}^{^{\prime}}$. Thus, returning to (\ref{proof_delta12prime}),
we get\vspace{-\belowdisplayskip}
\begin{equation}%
\begin{array}
[c]{ll}%
\delta_{12,j}^{^{\prime}}(\sigma_{k-1}^{+})= & x_{2,j}^{^{\prime}}%
(\sigma_{k-2})+\dot{x}_{2}(\sigma_{k-2})\sigma_{k-2,j}^{^{\prime}}\\
& -x_{2,j}^{^{\prime}}(\sigma_{k-1})-\dot{x}_{2}(\sigma_{k-1})\sigma
_{k-1,j}^{^{\prime}}%
\end{array}
\label{proof_delta2}%
\end{equation}
Recalling that $z_{12,j}^{^{\prime}}(\sigma_{k}^{+})=-\sigma_{k,j}^{^{\prime}%
}$ and combining (\ref{sigmaprime_k=0}),(\ref{proof_delta2}) into
(\ref{proof_sigma1}), we get\vspace{-\belowdisplayskip}
\begin{equation}%
\begin{array}
[c]{ll}%
\sigma_{k,j}^{^{\prime}} & =\sigma_{k-1,j}^{^{\prime}}+\frac{1}{{v_{1}}%
}[x_{2,j}^{^{\prime}}(\sigma_{k-2})+\dot{x}_{2}(\sigma_{k-2})\sigma
_{k-2,j}^{^{\prime}}\\
&
\begin{array}
[c]{l}%
-x_{2,j}^{^{\prime}}(\sigma_{k-1})-\dot{x}_{2}(\sigma_{k-1})\sigma
_{k-1,j}^{^{\prime}}]\\
=\sigma_{0,j}^{^{\prime}}+\frac{1}{{v_{1}}}[-x_{2,j}^{^{\prime}}(\sigma
_{k-1})-\dot{x}_{2}(\sigma_{k-1})\sigma_{k-1,j}^{^{\prime}}]
\end{array}
\end{array}
\label{sigmaprime_k>0}%
\end{equation}
where the last step follows from a recursive evaluation of $\sigma
_{k-1,j}^{^{\prime}}$ using (\ref{sigmaprime_k=0}) and (\ref{sigmaprime_k>0})
leading to many of the terms above canceling. This completes the proof.
$\blacksquare$

Let us now focus on event $J_{K}$ at time $\sigma_{K}$. It follows from the
reset condition in \eqref{x2dynamics} that\vspace{0pt}
\begin{equation}
{x}_{2,j}^{^{\prime}}(\sigma_{K}^{+})=\left\{
\begin{array}
[c]{l}%
{x}_{2,j}^{^{\prime}}(\sigma_{K})+{x}_{12,j}^{^{\prime}}(\sigma_{K})\\
+h_{2}(\sigma_{K}^{+})\sigma_{K,j}^{^{\prime}}\\
{x}_{2,j}^{^{\prime}}(\sigma_{K})+{x}_{12,j}^{^{\prime}}(\sigma_{K})
\end{array}
\right.
\begin{array}
[c]{l}%
\text{if }G_{2}(\sigma_{K})=1\\
\text{and }x_{2}(\sigma_{K})=0\\
\text{otherwise}\\
\end{array}
. \label{x2_prim}%
\end{equation}
Recall that $\delta_{12}(\sigma_{K}^{+})=0$ in \eqref{x2_prim}. If
$G_{2}(\sigma_{K})=1$ and $x_{2}(\sigma_{K})=0$, then $x_{2}(\sigma
_{K-1})-x_{2}(\sigma_{K})=0$, hence $x_{2}(\sigma_{K-1})=0$. It follows from
\eqref{x2dynamics} and \eqref{delta12_dynamics} that $\dot{x}_{2}(\sigma_{K-1})=0$.
Based on Case \textbf{1 }above, we get ${x}_{2,j}^{^{\prime}}(\sigma_{K-1}%
)=0$. Then, from Lemma 2, $\sigma_{K,j}^{^{\prime}}=\sigma_{0,j}^{^{\prime}}$
and \eqref{x2_prim} becomes\vspace{0pt}
\begin{equation}
{x}_{2,j}^{^{\prime}}(\sigma_{K}^{+})=\left\{
\begin{array}
[c]{l}%
{x}_{2,j}^{^{\prime}}(\sigma_{K})+{x}_{12,j}^{^{\prime}}(\sigma_{K})\\
+h_{2}(\sigma_{K}^{+})\sigma_{0,j}^{^{\prime}}\\
{x}_{2,j}^{^{\prime}}(\sigma_{K})+{x}_{12,j}^{^{\prime}}(\sigma_{K})
\end{array}
\right.
\begin{array}
[c]{l}%
\text{if }G_{2}(\sigma_{K})=1\\
\text{and }x_{2}(\sigma_{K})=0\\
\text{otherwise}\\
\end{array}
. \label{x2_prim2}%
\end{equation}
We conclude that the state derivative ${x}_{2,j}^{^{\prime}}(\sigma_{K}^{+})$
when event $S_{2}$ occurs is independent of all event time derivatives
$\sigma_{1,j}^{^{\prime}},\ldots,\sigma_{K,j}^{^{\prime}}$ and involves only
$\sigma_{0,j}^{^{\prime}}$, evaluated when the associated flow burst is initiated.

\textbf{(3)} \textit{Event }$G2R_{2}$: This is an endogenous event that occurs
when $g_{k}(x(\theta,t),\theta)=z_{2}(t)-\theta_{2}=0$. Based on
(\ref{eq:IPA_3}), $\tau_{k,j}^{^{\prime}}=\mathbf{1}[j=2]-z_{2,j}^{^{\prime}%
}(\tau_{k})$. Let $\rho_{k}$ be the last $R2G_{2}$ before $G2R_{2}$ occurs.
Applying (\ref{eq:IPA_1reduced}), we have $z_{2,j}^{^{\prime}}(\rho_{k}%
^{+})=z_{2,j}^{^{\prime}}(\tau_{k})$. and from (\ref{eq:IPA_2}) we get
$z_{2,j}^{^{\prime}}(\rho_{k}^{+})=-\rho_{k,j}^{^{\prime}}$. It follows that
$\tau_{k,j}^{^{\prime}}=\mathbf{1}[j=2]+\rho_{k,j}^{^{\prime}}$. Based on
(\ref{eq:IPA_2}), we have\vspace{-\belowdisplayskip}
\begin{equation}
{x}_{2,j}^{^{\prime}}(\tau_{k}^{+})=\left\{
\begin{array}
[c]{l}%
{x}_{2,j}^{^{\prime}}(\tau_{k})-h_{2}(\tau_{k})\tau_{k,j}^{^{\prime}}\\
{x}_{2,j}^{^{\prime}}(\tau_{k})
\end{array}
\right.
\begin{array}
[c]{l}%
\text{if }x_{2}(\tau_{k})>0\\
\text{otherwise}%
\end{array}
.
\end{equation}

\textbf{(4)} \textit{Event }$R2G_{2}$\textit{: }Let $\rho_{k}$ be the time of
this event and $\tau_{k}$ be the time of the last $G2R_{2}$ event before
$R2G_{2}$ occurs. Similar to \textbf{(3)} above, we get $\rho_{k,j}^{^{\prime
}}=\mathbf{1}[j=4]+\tau_{k,j}^{^{\prime}}$ and use this value in the
expression below which follows from (\ref{eq:IPA_2}):\vspace
{-\belowdisplayskip}
\begin{equation}
{x}_{2,j}^{^{\prime}}(\rho_{k}^{+})=\left\{
\begin{array}
[c]{l}%
{x}_{2,j}^{^{\prime}}(\rho_{k})+h_{2}(\tau_{k}^{+})\rho_{k,j}^{^{\prime}}\\
{x}_{2,j}^{^{\prime}}(\rho_{k})
\end{array}
\right.
\begin{array}
[c]{l}%
\text{if }x_{2}(\rho_{k})>0\\
\text{otherwise}%
\end{array}
.
\end{equation}

\textbf{(5)} \textit{Event }$J_{k}$: The analysis of this event has already
been done in Case \textbf{(2)} above, including Lemma 2\textbf{.}

\textbf{(6)} \textit{Event }$Z_{2}$: \ This is an endogenous event which is
triggered by $J_{k}$: if a traffic burst from node $1$ joins $x_{2}(t)$ at
$t=\tau_{k}$ and $x_{2}(\tau_{k}^{+})>\zeta_{2}$, this results in $Z_{2}$.
Since $r_{2}(\tau_{k}^{+})=1$ and $\dot{r}_{2}(t)=0$, we have $r_{2,j}
^{^{\prime}}(\tau_{k}^{+})=0$.

\textbf{(7)} \textit{Event }$\bar{Z}_{2}$: This is an endogenous event that
occurs when $g_{k}(x(\theta,t),\theta)=x_{2}(\theta,t)-\zeta_{2}=0$. Applying
(\ref{eq:IPA_3}), we have $\tau_{k,j}^{^{\prime}}=x_{2}^{^{\prime}}(\tau
_{k})/h_{2}(\tau_{k})$. Moreover, $r_{2}(\tau_{k}^{+})\equiv0$, therefore,
$r_{2,j}^{^{\prime}}(\tau_{k}^{+})+\dot{r}_{2}(\tau_{k}^{+})\tau_{k,j}^{+}=0$
and, since $\dot{r}_{2}(\tau_{k}^{+})=0$, we get $r_{2,j}^{^{\prime}}(\tau
_{k}^{+})=0$.

\subsubsection{\textit{IPA for Event Set}}
$\Phi_{12}=\{S_{12},E_{12},E_{1},G2R_{1},J_{k}\}\cup\{Z_{12}$,$\bar{Z}_{12}\}$

\textbf{(1)} \textit{Event }$S_{12}$: This event can be either exogenous or
endogenous. If $x_{1}(\tau_{k})>0$ or if $x_{1}(\tau_{k})=0,$ $\alpha
_{1}(t)>0$, $S_{12}$ is induced by event $R2G_{1}$ which is endogenous.
Otherwise, $S_{12}$ is exogenous event and occurs when $G_{1}(\tau_{k})=1$ and
$\alpha_{1}(\tau_{k})$ switches from zero to some positive value.

\textbf{Case (1a):} $S_{12}$\emph{ is induced by }$R2G_{1}$. Referring to our
analysis of $R2G_{1}$ (Case \textbf{(3)} for $\Phi_{1}$), we have already
evaluated $\tau_{k,j}^{^{\prime}}$. Then, applying (\ref{eq:IPA_2}), we
get\vspace{-\belowdisplayskip}
\begin{equation}
{x}_{12,j}^{^{\prime}}(\tau_{k}^{+})=\left\{
\begin{array}
[c]{l}%
{x}_{12,j}^{^{\prime}}(\tau_{k})-\alpha_{1}(\tau_{k}^{+})\tau_{k,j}^{^{\prime
}}\\
{x}_{12,j}^{^{\prime}}(\tau_{k})-h_{1}(\tau_{k}^{+})\tau_{k,j}^{^{\prime}}%
\end{array}
\right.
\begin{array}
[c]{l}%
\text{if }x_{1}(\tau_{k})=0 \text{ and }\\
0<\alpha_{1}(\tau_{k})\leq\beta_{1}(\tau_{k})\\
\text{otherwise}%
\end{array}
.
\end{equation}

\textbf{Case(1b)} $S_{12}$\emph{ is exogenous}. In this case, $\tau
_{k,j}^{^{\prime}}=0$ and applying (\ref{eq:IPA_2}) gives ${x}_{12,j}
^{^{\prime}}(\tau_{k}^{+})={x}_{12,j}^{^{\prime}}(\tau_{k})$.

\textbf{(2)} \textit{Event }$E_{12}$: This event occurs when the traffic burst
in queue $12$ joins queue $2$. This is an endogenous event that occurs when
$g_{k}(x(\theta,\tau_{k}),\theta)=z_{12}(\tau_{k})-\delta_{12}(\tau_{k})=0$
and $\delta_{12}(\tau_{k}^{+})=0$. When this happens, it follows from the
reset condition in (\ref{x12_dynamics}) that $x_{12,j}^{^{\prime}}(\tau
_{k}^{+})=0$.

\textbf{(3)} \textit{Event }$E_{1}$: This is an endogenous event that occurs
when $g_{k}(x(\theta,t),\theta)=x_{1}(t)=0$. Applying (\ref{eq:IPA_3}), we get
$\tau_{k,j}^{^{\prime}}=-\frac{{x}_{1,j}^{^{\prime}}(\tau_{k})}{\alpha
_{1}(\tau_{k})-h_{1}(\tau_{k})}$. Thus, using (\ref{eq:IPA_2}), we
get\vspace{-0.6pt}
\begin{equation}%
\begin{array}
[c]{ll}%
{x}^{^{\prime}}_{12,j}(\tau^{+}_{k}) & ={x}^{^{\prime}}_{12,j}(\tau
_{k})+(h_{1}(\tau_{k})-\alpha_{1}(\tau_{k})) \tau^{^{\prime}}_{k,j}\\
& ={x}^{^{\prime}}_{12,j}(\tau_{k})+{x}^{^{\prime}}_{1,j}(\tau_{k})
\end{array}
.
\end{equation}

\textbf{(4)} \textit{Event }$G2R_{1}$: This is an endogenous event that occurs
when $g_{k}(x(\theta,t),\theta)=z_{1}(t)-\theta_{1}=0$. It was shown under the
analysis for events in $\Phi_{1}$ that for $G2R_{1}$ we have $\tau
_{k,j}^{^{\prime}}=\mathbf{1}[j=i]+\rho_{k,j}^{^{\prime}}$ where $\rho_{k}$ is
the time of the last $R2G_{1}$ event before $G2R_{1}$ occurs. Using this
value, we can the evaluate the following which follows from (\ref{eq:IPA_2}
):\vspace{-\belowdisplayskip}
\begin{equation}
{x}_{12,j}^{^{\prime}}(\tau_{k}^{+})=\left\{
\begin{array}
[c]{l}%
{x}_{12,j}^{^{\prime}}(\tau_{k})+\alpha_{1}(\tau_{k})\tau_{k,j}^{^{\prime}}\\
\\
{x}_{12,j}^{^{\prime}}(\tau_{k})+h_{1}(\tau_{k})\tau_{k,j}^{^{\prime}}%
\end{array}
\right.
\begin{array}
[c]{l}%
\text{if }x_{1}(\tau_{k})=0\\
\text{and }\alpha_{1}(t)\leq\beta_{1}(t)\\
\text{otherwise}\\
\end{array}
\end{equation}

\textbf{(5)} \textit{Event }$J_{k}$: The analysis of this event has already
been done in Case \textbf{(2)} above, including Lemma 2\textbf{.}

\textbf{(6)} \textit{Event }$Z_{12}$: This is an endogenous event that occurs
when $g_{k}(x(\theta,t),\theta)=x_{12}(\theta,t)-\zeta_{12}=0$. Applying
(\ref{eq:IPA_3}), we have\vspace{-\belowdisplayskip}
\[
\tau_{k,j}^{^{\prime}}=\left\{
\begin{array}
[c]{l}%
-\frac{x_{12,j}^{^{\prime}}(\tau_{k})}{\alpha_{1}(\tau_{k})}\\
\\
-\frac{x_{12,j}^{^{\prime}}(\tau_{k})}{h_{1}(\tau_{k})}%
\end{array}
\right.
\begin{array}
[c]{l}%
\text{if }x_{1}(\tau_{k})=0\\
\text{and }\alpha_{1}(t)\leq\beta_{1}(t)\\
\text{otherwise}%
\end{array}
.
\]
Since $r_{12}(\tau_{k}^{+})=1$ and $\dot{r}_{12}(t)=0$, we have $r_{12,j}
^{^{\prime}}(\tau_{k}^{+})=0$.

\textbf{(7)} \textit{Event }$\bar{Z}_{12}$: This is triggered by event
$E_{12}$ when the traffic burst in queue $12$ joins queue $2$ and we reset
$x_{12}(\tau_{k}^{+})=0$. Since $r_{12}(\tau_{k}^{+})=0$ and $\dot{r}
_{12}(t)=0$, we have $r_{12,j}^{^{\prime}}(\tau_{k}^{+})=0$. \vspace
*{-\baselineskip}

\subsection{Cost Function Derivatives}

Returning to (\ref{ave_cost}), (\ref{power_cost1}), and (\ref{th_cost1}),
recall that the IPA estimator consists of the gradient formed by the sample
performance derivatives $\frac{dF}{d\theta_{j}}$, which in turn depend on the
state derivatives that we have evaluated in the previous section. The
derivation of the IPA estimator for the Average Queue cost function in
(\ref{ave_cost}) is similar to that in \cite{geng2012traffic} and related
prior work and is omitted. Instead, we concentrate on the two new cost
functions (\ref{power_cost1}), and (\ref{th_cost1}).

For the Power cost function, we derive $\frac{dF_{i,m}(\theta)}{d\theta_{j}}$
from (\ref{power_cost2}), from which $\frac{dF}{d\theta_{j}}$ is obtained by
adding over all $M_{i}$ NEPs of each queue $i$ over $[0,T]$:\vspace
*{-\belowdisplayskip}
\[%
\begin{array}
[c]{ll}%
\frac{dF_{i,m}(\theta)}{d\theta_{j}} & =Px_{i,j}^{^{\prime}}(\theta
,t)\int_{\xi_{i,m}(\theta)}^{\eta_{i,m}(\theta)}w_{i}x_{i}^{P-1}(\theta,t)dt\\
& =P[x_{i,j}^{^{\prime}}(\xi_{i,m}^{+})\int_{\xi_{i,m}(\theta)}^{t_{i,m}^{1}
}w_{i}x_{i}^{P-1}(\theta,t)dt\\
& +\sum_{j=2}^{J_{i,m}}x_{i,j}^{^{\prime}}((t_{i,m}^{j})^{+})\int
_{t_{i,m}^{j-1}}^{t_{i,m}^{j}}w_{i}x_{i}^{P-1}(\theta,t)dt\\
& +x_{i,j}^{^{\prime}}((t_{i,m}^{J_{i,m}})^{+})\int_{t_{i,m}^{J_{i,m}}}
^{\eta_{i,m}}w_{i}x_{i}^{P-1}(\theta,t)dt],
\end{array}
\vspace*{-\belowdisplayskip}
\]
where $t_{i,m}^{j},j=1,...,J_{i,m}$ is the occurrence time of the $j$th event
in the $m$th NEP of queue $i$. The state derivative is determined on an
event-driven basis using $x_{i,j}^{^{\prime}}(\tau_{k}^{+})$ corresponding to
the event occurring at time $\tau_{k}$; for instance, if $G2R_{1}$ occurs at
node $1$, then (\ref{state_derivative_G2R_i}) is invoked with $i=1$.

For the Threshold cost function, we know that $r^{^{\prime}}(\theta,t)=0$ and
it follows from (\ref{th_cost2}):\vspace*{-\belowdisplayskip}
\[%
\begin{array}
[c]{ll}%
\frac{dF_{i,m}(\theta)}{d\theta_{j}} & =\int_{\gamma_{i,m}(\theta)}
^{\psi_{i,m}(\theta)}w_{i}r_{i,j}^{^{\prime}}(\theta,t)dt-w_{i}r_{i}
(\theta,\gamma_{i,m}^{+})\gamma_{i,m,j}^{^{\prime}}\\
& +w_{i}r_{i}(\theta,\psi_{i,m}^{-})\psi_{i,m,j}^{^{\prime}}\\
& =w_{i}(\psi_{i,m,j}^{^{\prime}}-\gamma_{i,m,j}^{^{\prime}}),
\end{array}
\]
Note that in this case the derivative depends only on $\psi_{i,m,j}^{^{\prime
}},$ $\gamma_{i,m,j}^{^{\prime}}$, the event time derivatives in
(\ref{eventtime_derivative_B_i}),(\ref{eventtime_derivative_EB_i}) for
$i=1,3,4$ and the corresponding event time derivatives in Cases
(\textbf{6),(7)} for each of sets $\Phi_{22}$ and $\Phi_{12}$.

\section{SIMULATION RESULTS}

In this section, we use the derived IPA estimators in order to optimize the
green light cycles in the two-intersection model of Fig.
\ref{Twotrafficintersections}. We stress that this model is simulated as a
Discrete Event System (DES) with individual vehicles rather than flows, so
that the resulting estimators are based on actual observed data. This is made
possible by the fact that all SFM events in the sets $\Phi_{i}$,
$i=1,\ldots,4$, and $\Phi_{12}$ coincide with those of the DES, therefore they
are directly observable along with their occurrence times.

We assume that all vehicle arrival processes are Poisson (recall, however,
that IPA is \emph{independent of these distributions}) with rates $\bar
{\alpha_{i}},i=1,3,4$, and that the vehicle departure rate $h_{i}(t)$ on each
non-empty road is constant. In \cite{geng2015multi}, only one controllable
parameter per intersection was considered by setting $\theta_{i}+\theta
_{\bar{\imath}}=C$. Here, we relax this constraint. Moreover, we limit each
controllable parameter so that $\theta_{i}\in\lbrack\theta_{i,min}%
,\theta_{i,max}]$. In our simulations, $\alpha_{i}(\tau_{k})$ is estimated
through $N_{a}/t_{w}$ by counting the number of arriving vehicles $N_{a}$ over
a time interval $[0,t_{w}]$ and $h_{i}(t)$ is estimated using the same method
as in \cite{fleck2016adaptive}. Three sets of simulations are presented below,
one for each of the three cost metrics in \eqref{ave_cost},
\eqref{power_cost1} and \eqref{th_cost1}.

\begin{figure}[ptb]
\centering
\includegraphics[scale=0.067]{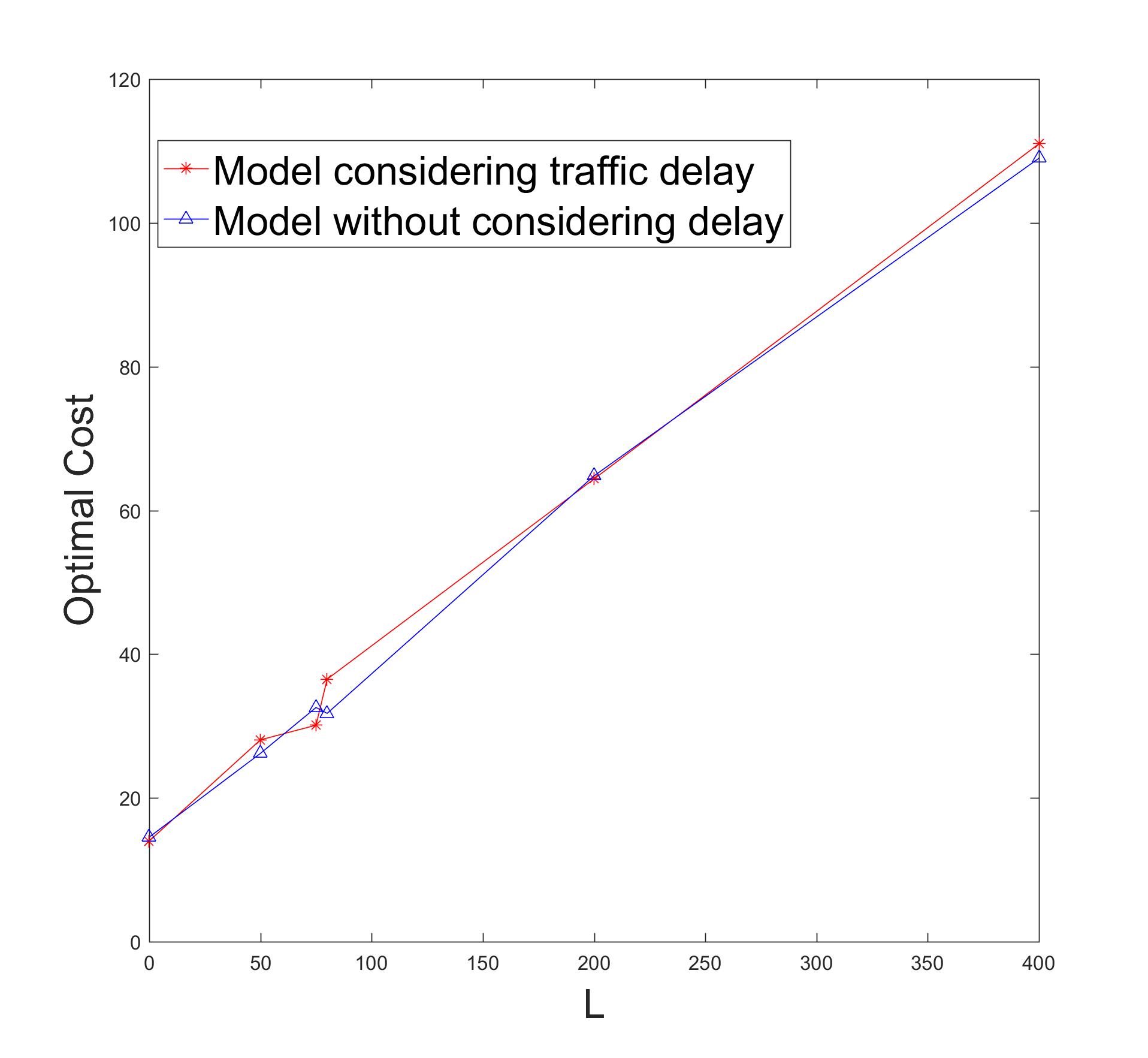} \caption{Comparison of Optimal
Average Queue Cost vs L.}%
\label{compare_vs_L}%
\end{figure}

\textbf{1. Average Queue Cost Function}. We minimize metric \eqref{ave_cost},
over $[0,T]$. All three arrival processes are Poisson with rates $\bar{\alpha
}=[0.41,0.45,0.32]$ and the departure rates at roads $1,2,3,4$ are
$[1.2,1.3,1.2,1.1]$. We choose $T=1000$s, $w_{i}=1$ and $\theta_{i}\in
\lbrack10,50]$ for all $i$, and the initial $\theta_{i}$ values are
$[40,20,20,40]$. Figure \ref{compare_vs_L} shows the optimal cost (averaged
over $10$ sample paths) considering the transit delay in SFM between
intersections (red curve) and ignoring this delay (blue curve) as a function
of $L$. In this case, delay has no effect on the long term total average queue
length, as expected. However, this metric may not accurately capture traffic congestion.

\begin{figure}[ptb]
\centering
\includegraphics[scale=0.19]{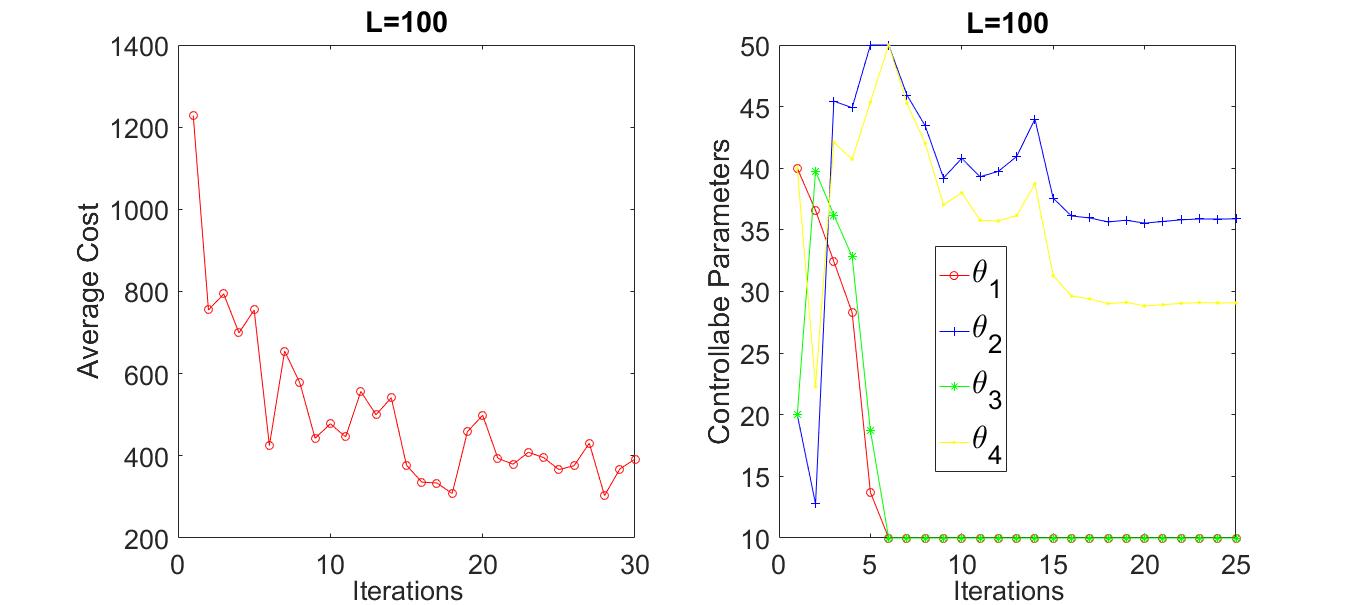} \caption{Optimal Power
Cost Function vs Iterations.}%
\label{costx^2_vs_iteration}%
\end{figure}

\begin{figure}[ptb]
\vspace*{-\belowdisplayskip} \centering
\includegraphics[scale=0.07]{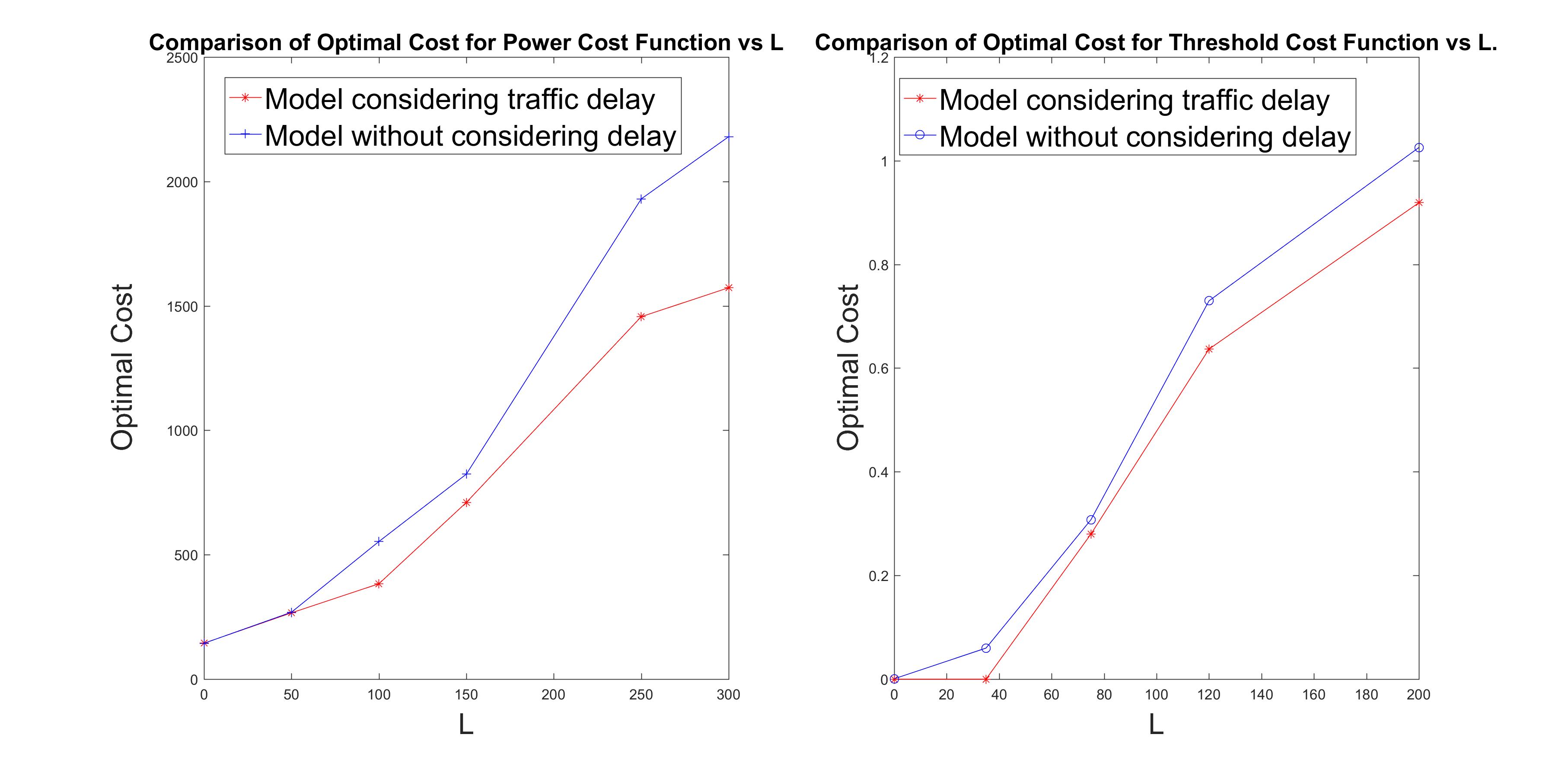} \caption{Comparison of
Optimal Cost with/without delay vs $L$.}%
\label{merge_x_2_th_compare}%
\end{figure}

\textbf{2. Power Cost Function, }$P=2$. For the same settings as before and a
quadratic queuing cost, Fig. \ref{costx^2_vs_iteration} shows how this cost
function and the associated controllable parameters converge when $L=100$,
achieving a $40\%$ cost decrease. In the left plot of Fig.
\ref{merge_x_2_th_compare}, we use the SFM both including the transit delay
and ignoring this delay in order compare the optimal costs under these two
models. Clearly, including delays in our IPA estimators for $L>0$ achieves a
lower cost, with the gap increasing as $L$ increases.



\begin{figure}[ptb]
\centering
\includegraphics[scale=0.19]{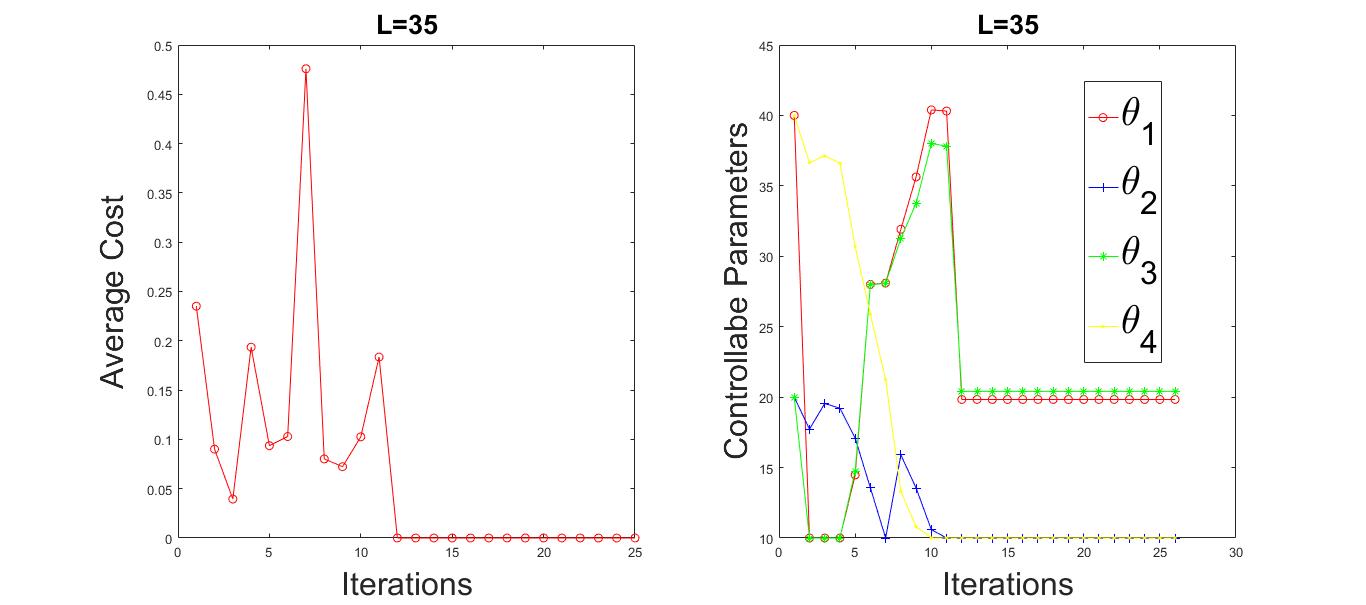} \caption{Optimal
Threshold Cost Function vs Iterations.}%
\label{cost_th_vs_iteration}%
\end{figure}

\textbf{3. Threshold Cost Function}. For the same settings and a common
threshold $\zeta_{i}=25$ for all $i$ and with $L=35$, Fig.
\ref{cost_th_vs_iteration} shows how this cost function and the associate
controllable parameters converge, with the cost converging to its zero lower
bound, therefore, in this case we see that our approach reaches the global
optimum. In the right plot of Fig. \ref{merge_x_2_th_compare}, we apply the
SFM considering both the transit delay between intersections and ignoring this
delay so as to compare the resulting optimal costs.Once again, including
delays achieves a lower cost, with the gap increasing as $L$ increases.

\begin{figure}[ptb]
\centering
\includegraphics[scale=0.49]{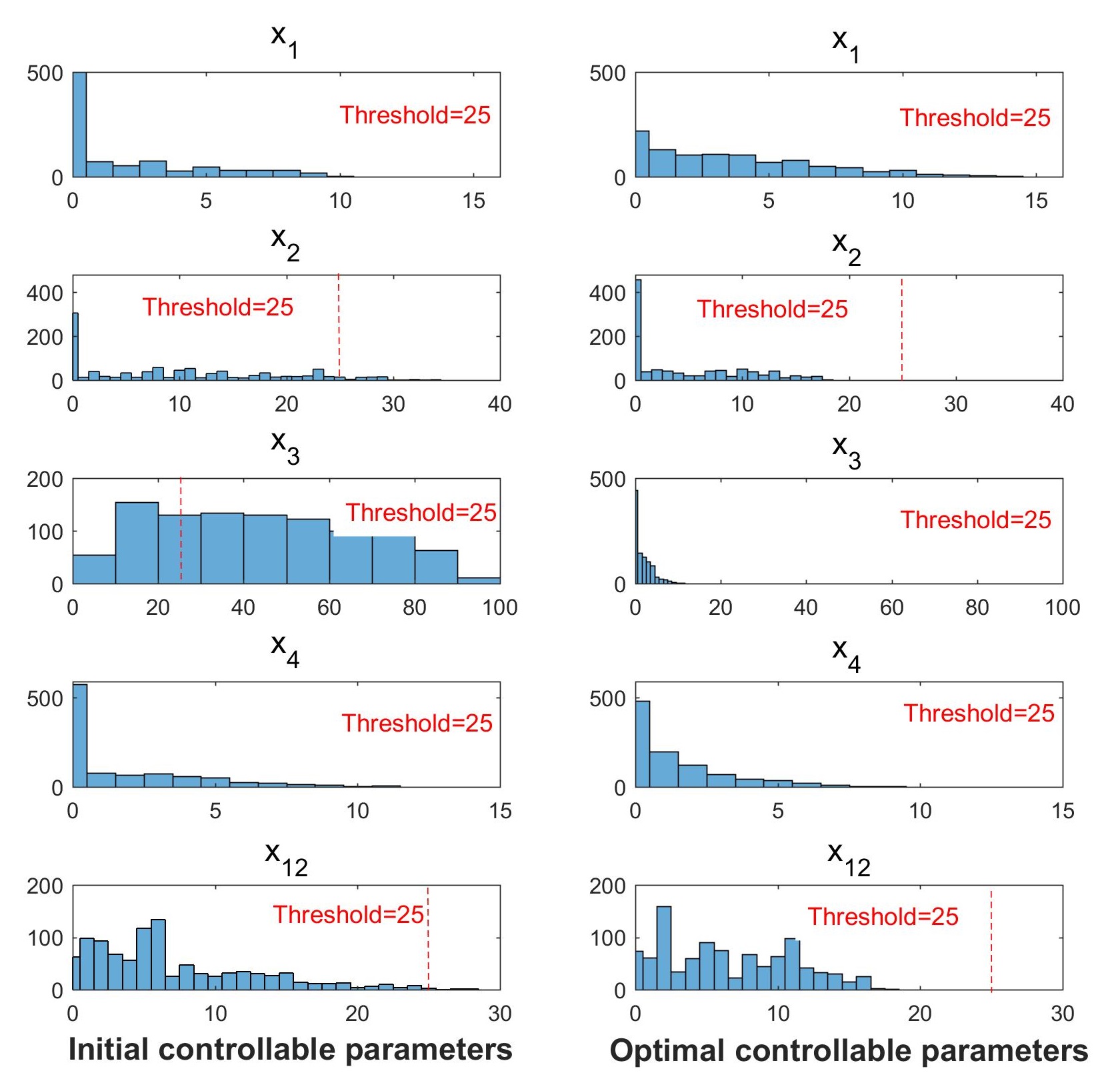} \caption{Distribution of
queue lengths under $L=35$.}%
\label{L=35_hiscompare}%
\end{figure}In Fig. \ref{L=35_hiscompare}, we provide histograms of the queue
contents when $L=35$. On the left, the controllable parameters are at their
initial values $[40,20,20,40]$ and we can see that queues $2$, $3$, and $12$
frequently exceed the threshold. Under the optimal solution we obtain (right
side) taking the transit delay between intersections into account, observe
that no queue ever exceeds the threshold over $[0,T]$, hence the optimal cost
$0$ is obtained. Moreover, note that the probabilities that $x_{2}(t)=0$ and
$x_{3}(t)=0$ significantly increase indicating a much improved traffic balance.

\section{CONCLUSIONS AND FUTURE WORK}

We have extended SFMs to allow for delays which can arise in the flow
movement. We have applied this framework to the multi-intersection traffic
light control problem by including transit delays for vehicles moving from one
intersection to the next and developed IPA for this extended SFM in order to
derive on-line gradient estimates of several congestion cost metrics with
respect to the controllable green/red cycle lengths, including two new cost
metrics that better capture congestion. Our simulation results show that the
inclusion of delays in our analysis leads to improved performance relative to
models that ignore delays. Future work aims at extensions to allow traffic
blocking between intersections and allowing multiple traffic bursts between intersections.

\vspace*{-\baselineskip}

\bibliographystyle{plain}
\bibliography{TLCbib2}

\end{document}